\documentclass[preprint]{elsarticle}
\usepackage[utf8]{inputenc}
\usepackage{natbib}
\usepackage{amsmath}
\usepackage{amsfonts}
\usepackage{amssymb}
\usepackage{graphicx}
\usepackage{tabularx}
\usepackage{booktabs}
\usepackage{siunitx}
\usepackage{wrapfig}
\usepackage[version=3]{mhchem}
\usepackage{subfigure}
\usepackage{float}
\usepackage{multirow}
\usepackage{caption}
\usepackage{xcolor}
\biboptions{sort&compress}	
\usepackage{fancyhdr}
\newenvironment{keywords}%
   {\begin{trivlist}\item[]{\bfseries\sffamily Keywords:}\ }
   {\end{trivlist}}

\journal{Elsvier, License CC-BY-NC-ND 4.0}
\title{Intake Design for an Atmosphere-Breathing Electric Propulsion System (ABEP)}

\author[irs]{F.~Romano\corref{cor1}\fnref{fn1}}
\ead{romano@irs.uni-stuttgart.de}

\author[irs]{J.~Espinosa-Orozco\corref{cor1}\fnref{fn2}}
\ead{jesuseo@kth.se}

\author[irs]{M.~Pfeiffer\corref{cor1}\fnref{fn3}}

\author[irs]{G.~Herdrich\fnref{fn4}}
\author[UniMAN]{N.H.~Crisp}

\author[UniMAN]{P.C.E.~Roberts}
\author[UniMAN]{B.~E.A.~Holmes}
\author[UniMAN]{S.~Edmondson}
\author[UniMAN]{S.~Haigh}
\author[UniMAN]{S.~Livadiotti}
\author[UniMAN]{A.~Macario-Rojas}
\author[UniMAN]{V.T.~A.~Oiko}
\author[UniMAN]{L.A.~Sinpetru}
\author[UniMAN]{K.~Smith}

\author[Deimos]{J.~Becedas}
\author[Deimos]{V.~Sulliotti-Linner}

\author[GomSpace]{M.~Bisgaard}
\author[GomSpace]{S.~Christensen}
\author[GomSpace]{V.~Hanessian}
\author[GomSpace]{T.~Kauffman Jensen}
\author[GomSpace]{J.~Nielsen}

\author[irs]{Y.-A.~Chan}
\author[irs]{S.~Fasoulas}
\author[irs]{C.~Traub}

\author[UPC]{D.~García-Almiñana}
\author[UPC]{S.~Rodríguez-Donaire}
\author[UPC]{M.~Sureda}

\author[UCL]{D.~Kataria}

\author[Eurocons]{B.~Belkouchi}
\author[Eurocons]{A.~Conte}
\author[Eurocons]{S.~Seminari}
\author[Eurocons]{R.~Villain}

\fntext[fn1]{Research Associate, romano@irs.uni-stuttgart.de}
\fntext[fn2]{Research Guest, jesuseo@kth.se}
\fntext[fn3]{Head of Numerical Modelling and Simulations, mpfeiffer@irs.uni-stuttgart.de}
\fntext[fn4]{Head Plasma Wind Tunnels and Electric Propulsion, herdrich@irs.uni-stuttgart.de}

\address[irs]{Institute of Space Systems (IRS), University of Stuttgart, 70569, Germany}
\address[UniMAN]{The University of Manchester, Oxford Road, Manchester, M13 9PL, UK}
\address[Deimos]{Elecnor Deimos Satellite Systems, C/ Francia 9, 13500, Puertollano, Spain}
\address[GomSpace]{GomSpace AS, Langagervej 6, Aalborg East 9220, Denmark}
\address[UPC]{UPC-BarcelonaTECH, Colom 11, TR5 - 08222 Terrassa, Spain}
\address[UCL]{Mullard Space Science Laboratory, University College London, Holmbury St.~Mary, Dorking, Surrey, RH5 6NT, UK}
\address[Eurocons]{Euroconsult, 86 Boulevard de Sébastopol, 75003 Paris, France}
\begin{document}

\begin{abstract}
Challenging space missions include those at very low altitudes, where the atmosphere is source of aerodynamic drag on the spacecraft. To extend the lifetime of such missions, an efficient propulsion system is required. One solution is Atmosphere-Breathing Electric Propulsion (ABEP) that collects atmospheric particles to be used as propellant for an electric thruster. The system would minimize the requirement of limited propellant availability and can also be applied to any planetary body with atmosphere, enabling new missions at low altitude ranges for longer times. IRS is developing, within the H2020 DISCOVERER project, an intake and a thruster for an ABEP system. The article describes the design and simulation of the intake, optimized to feed the radio frequency (RF) Helicon-based plasma thruster developed at IRS. The article deals in particular with the design of intakes based on diffuse and specular reflecting materials, which are analysed by the PICLas DSMC-PIC tool. Orbital altitudes $h=150-\SI{250}{\kilo\meter}$ and the respective species based on the NRLMSISE-00 model (\ce{O},~\ce{N2},~\ce{O2},~\ce{He},~\ce{Ar},~\ce{H},~\ce{N}) are investigated for several concepts based on fully diffuse and specular scattering, including hybrid designs. The major focus has been on the intake efficiency defined as $\eta_c=\dot{N}_{out}/\dot{N}_{in}$, with $\dot{N}_{in}$ the incoming particle flux, and $\dot{N}_{out}$ the one collected by the intake. Finally, two concepts are selected and presented providing the best expected performance for the operation with the selected thruster. The first one is based on fully diffuse accommodation yielding to $\eta_c<0.46$ and the second one based un fully specular accommodation yielding to $\eta_c<0.94$. Finally, also the influence of misalignment with the flow is analysed, highlighting a strong dependence of $\eta_c$ in the diffuse-based intake while, for the specular-based intake, this is much lower finally leading to a more resilient design while also relaxing requirements of pointing accuracy for the spacecraft.
\end{abstract}
 \maketitle

\section*{Nomenclature}
\noindent
ABEP: Atmosphere-Breathing Electric Propulsion\\
DSMC: Direct Simulation Monte Carlo \\
FMF: Free Molecular Flow \\
GSI: Gas-Surface Interaction \\
IPT: RF Helicon-based Plasma Thruster\\
VLEO: Very Low Earth Orbit\\
SC: Spacecraft 

\section{Introduction} 
\subsection{ABEP Concept}
\begin{figure}[H]
	\centering
	\includegraphics[width=12.5cm]{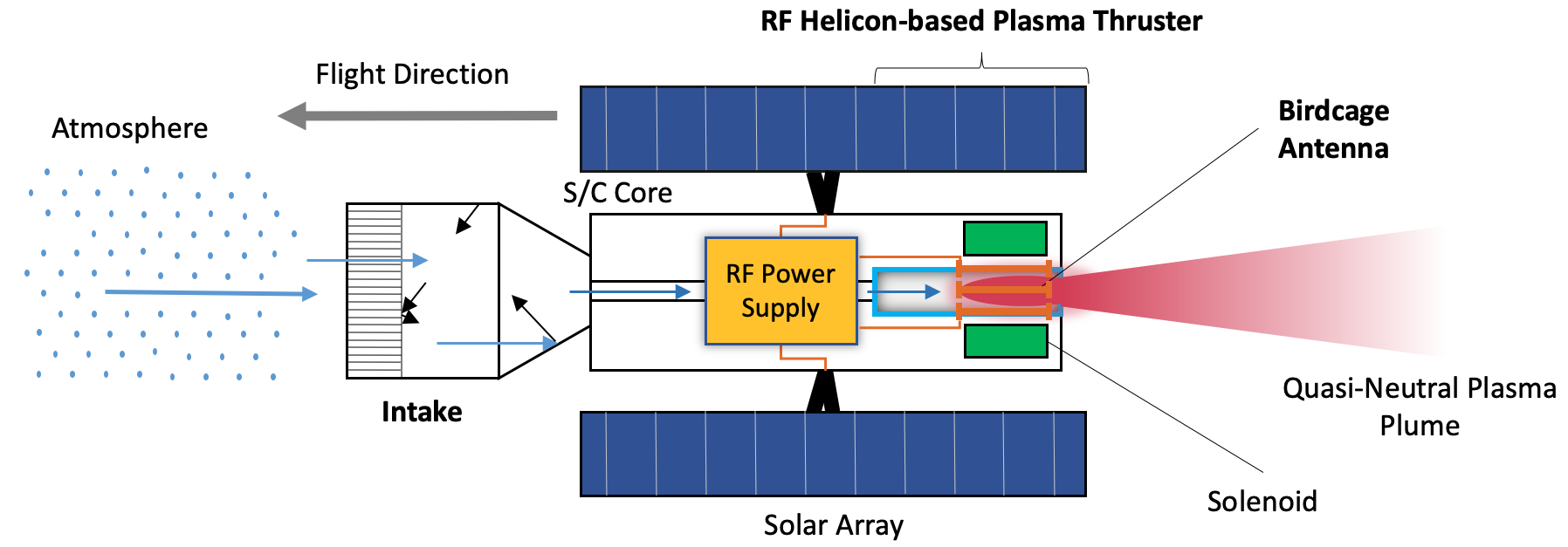}
	\caption{Atmosphere-Breathing Electric Propulsion Concept using the RF Helicon-based plasma thruster~\cite{romanoacta2}.}
	\label{fig:ABEP}
\end{figure}

An atmosphere-breathing electric propulsion system (ABEP), see Fig.~\ref{fig:ABEP}, is composed of two main components: the intake and the electric thruster. The system is designed for satellites orbiting at very low altitude altitudes, for example in very low Earth orbit (VLEO), defined for altitudes $h<~\SI{450}{\kilo\meter}$~\cite{vleobenefit}. The ABEP system collects the residual atmospheric particles encountered by the satellite through the intake and uses them as propellant for the electric thruster. The system is theoretically applicable to any planetary body with an atmosphere, and can drastically reduce the on-board propellant storage requirement while extending the mission's lifetime~\cite{romanoacta}.  Many ABEP concepts have been investigated in the past based on radio frequency ion thrusters (RIT)~\cite{di2007ram,presitael1,presitael2}, ECR-based thruster~\cite{JAXA,JAXA2,JAXA3,JAXA4,JAXA5}, Hall-effect thrusters (HET)~\cite{busek,busek2,SITAEL2015,SITAEL2016,SITAEL2017,SITAEL2019a,SITAEL2019b}, and plasma thrusters~\cite{shabshelowitz2013study}. The only laboratory tested ABEP systems to date are the ABIE, composed of an annular intake and an ECR-based thruster into one device~\cite{JAXA,JAXA2,JAXA3,JAXA4}, and the RAM-HET system, comprised of the intake and a HET assembled into one device~\cite{SITAEL2015,SITAEL2016,SITAEL2017,SITAEL2019a,SITAEL2019b}. 

\subsection{Intake Working Principle}
\begin{figure}[H]
	\centering
	\includegraphics[width=10cm]{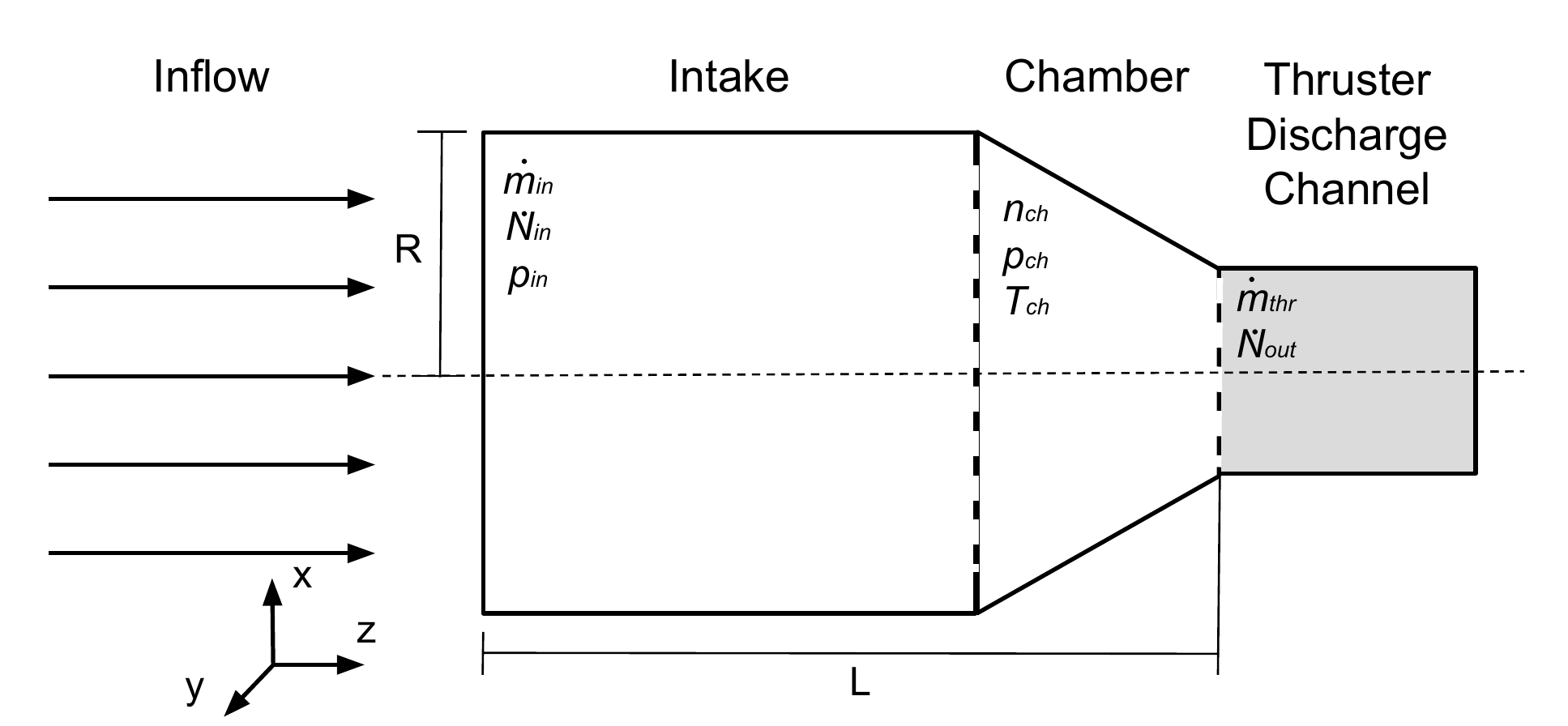}
	\caption{Intake General Design including the Thruster's Discharge Channel.}
	\label{fig:Gen_Diag}
\end{figure}

The intake is the device of an ABEP system that collects and delivers the atmospheric particles to the electric thruster, the general design is shown in Fig.~\ref{fig:Gen_Diag}. The intake must ensure efficient collection of atmospheric particles and provide the required density, pressure, and mass flow for the thruster’s operation, while minimizing the required frontal area.

The number density is $n$ and $\dot{N}$ is the particle flux. 
A key parameter to evaluate the intake performance, is the collection efficiency $\eta_c(h)$ defined in Eq.~\ref{Eq:intake_eff}, where $\dot{N}_{out}$ is the collected particle flux and $\dot{N}_{in}$ is the incoming one, the subscript $i$ refers to each different particle species $i= [1, N_s]$, where $N_s$ is the total number of species found in the inflow environment. As the VLEO composition is highly variable, the task to efficiently collect each species in a variable density range is challenging.

\begin{equation}
 \eta_c (h) =  \frac{\sum_{i=1}^{N_s}\dot{N}_{out_i}}{\sum_{i=1}^{N_s}\dot{N}_{in_i}}
 \label{Eq:intake_eff}
\end{equation}
A further key parameter is the pressure at the intake chamber section $p_{ch}$, which is an input for the thruster. By applying the ideal gas assumption $p_{ch}$ is derived as in Eq.~\ref{Eq:pres_equ1}, where $n_{ch}$ and $T_{ch}$ are the particle number density and temperature at the chamber section, and $k_b= \SI{1.380649E-23}{\joule\per{\kelvin}}$ is the Boltzmann constant.

\begin{equation}
p_{ch} = \sum_{i=1}^{N_s} n_{ch_i}~ k_b~ T_{ch}
\label{Eq:pres_equ1}
\end{equation}

Finally, the output mass flow to the thruster $\dot{m}_{thr}$ is obtained as in Eq.~\ref{Eq:mass_flow}, where $m_{p}$ is the particle mass.

\begin{equation}
 \dot{m}_{thr}(h) = \sum_{i=1}^{N_s} m_{p_i} ~ \dot{N}_{out_i} (h)
 \label{Eq:mass_flow}
\end{equation}

Within this study, hexagonal and circular geometries, for the whole intake but also for the ducts are investigated. To differentiate between them, the aspect ratio ($AR$) is defined in Eq.~\ref{Eq:AR} and used, where $S$ is the hexagonal duct short diagonal, $D=2R$ is the circle diameter, and $L$ the intake length.

\begin{equation}
 AR = \frac{L}{R} ~~~ \Bigg\{ \begin{array}{l}  R = \frac{S}{2} ~~~ \text{for the hexagonal-shape based intake} \\ \\
                                                R = \frac{D}{2} ~~~ \text{for the parabolic-shape based intake} 
                                				  \end{array}
\label{Eq:AR}
\end{equation}

\subsection{Intake Literature Review}
Over the last decades, several research groups have approached the intake design for an ABEP system~\cite{SINGH201515}. Most of those are based on diffuse scattering materials, and implement a front section of small ducts to reduce the backflow, acting as molecular trap, to achieve higher $\eta_c < 0.5$. The intake designed by Nishiyama and JAXA~\cite{JAXA,JAXA2,JAXA3,JAXA4,JAXA5}, uses long ducts packed in a ring-shaped section surrounding the SC body and ending on a conical reflector at the back, $\eta_c<0.46$. There they are ionized by electron cyclotron resonance (ECR) and accelerated by grids to produce thrust. Busek~Co.~Inc.~\cite{busek,busek2} designed the MABHET spacecraft for Mars orbit equipped with an ABEP system based on HET technology. It has a long and slender cylindrical intake in front of the SC with a honeycomb duct section at the front to operate as a molecular trap delivering $\eta_c=0.2-0.4$. The intake designs developed at IRS~\cite{romanoiepc, romanosp2016,tilman} are based on a long slender cylindrical intake with a honeycomb duct section in the front optimized for both Earth and Mars atmosphere that can achieve $\eta_c=0.43$. The Central Aerohydrodynamic Institute TsAGI~\cite{TSAGI1,TSAGI2,filatyev2019control,TSaGI2018a,TSAGI2018ab,erofeev2017air,kanev2015electro} developed a similar concept, but the honeycomb is changed for a squared duct section delivering $\eta_c=0.33-0.34$. SITAEL and the von Karman Institute for Fluid Dynamics, Aeronautics and Aerospace~\cite{SITAEL2015,SITAEL2016,SITAEL2017,SITAEL2019a,SITAEL2019b,VKI}, refined the intake design that started with the 2007 ESA RAM-EP study~\citep{di2007ram}. The intake features several coaxial cylinders connected by plates which form divisions of different AR between each cylinder delivering $\eta_c=0.28-0.32$. This has been tested in combination with a modified  HET for ABEP operation and is the only ABEP system of intake and thruster in one device that has been tested for continuos operation, to date, in the laboratory~\cite{SITAEL2015,SITAEL2016,SITAEL2017,SITAEL2019a,SITAEL2019b,VKI}.
The Lanzhou Institute of Space Technology and Physics~\cite{Intake_YANWU2015}, designed an active intake with a two stage system, a passive multi-hole plate followed by an active compression with a turbo pump system that can achieve $\eta_c=0.42-0.58$. Finally, the University of Colorado~\cite{IntakeParabola_JACKSON2018} investigated specular scattering and finally designed an intake using the optical proprieties of a parabola leading to an estimated $\eta_c > 0.9$. A summary of these intake studies and their main features and performances are shown in Tab.~\ref{tab:intakes}.

\begin{table}[h]
\caption{Literature Intake Main Performances.}
\label{tab:intakes}
\begin{tabular}{lcccc}
\toprule
Design & Act./Pass. & Scattering & Ducts & $\eta_c$,~- \\
\midrule
ABIE~\cite{JAXA,JAXA2,JAXA3,JAXA4,JAXA5} & P & Diffuse & Yes & $0.28-0.46$ \\
BUSEK~\cite{busek,busek2} & P &  Diffuse & Yes & $0.20-0.40$\\
U.~Stuttgart IRS~\cite{romanoiepc, romanosp2016}  & P &  Diffuse & Yes & $0.43$ \\
SITAEL,~VKI~\cite{SITAEL2015,SITAEL2016,SITAEL2017,SITAEL2019a,SITAEL2019b,VKI} & P &  Diffuse & Yes & $0.28-0.32$ \\
Lanzhou Inst.~\cite{Intake_YANWU2015} & A &  Diffuse & Yes & $0.42-0.58$ \\
TsAGI/RIAME~\cite{TSAGI1,TSAGI2,filatyev2019control,TSaGI2018a,TSAGI2018ab,erofeev2017air,kanev2015electro} & P & Diffuse & Yes & $0.33-0.34$ \\
U.~Colorado~\cite{IntakeParabola_JACKSON2018} & P &  Specular & No & $<0.90$ \\
\bottomrule
\end{tabular}
\end{table}

Finally, each intake design always needs to be tailored based on the thruster applied and its requirements and also on the mission profile: expected operational orbits (altitudes and inclinations) and duration (i.e. to account for solar activity variation).

\section{Gas-Surface Interactions (GSI)}
Rarefied flows are described by the gas-kinetic Boltzmann equation, which is commonly solved numerically using the DSMC method~\cite{bird1994molecular}. Due to the few collisions that occur in rarefied flows, flow parameters are typically strongly dominated by wall interactions of the particles, in particular in VLEO~\cite{LIVADIOTTI2020}. Therefore, modelling the surface interactions of the particles as accurately as possible plays an important role in this flow regime and thus also in the design of the intake~\cite{PInt_Diez2013,PInt_LIANG2018}. In general, particles can either be absorbed, scattered, or undergo a chemical reaction at surfaces. Accurate modelling of these processes is highly complex due to the lack of sufficient knowledge of surface properties, such as finish, cleanliness, adsorbed gas layers, and especially when placed in VLEO environment and the interactions that occur with different gases, particularly atomic oxygen (AO). Finally, due to the requirements for the intake design, the Maxwell model is introduced.

\subsection{Maxwell Model}
\begin{figure}[H]
 \centering
 \includegraphics[width=.9\textwidth]{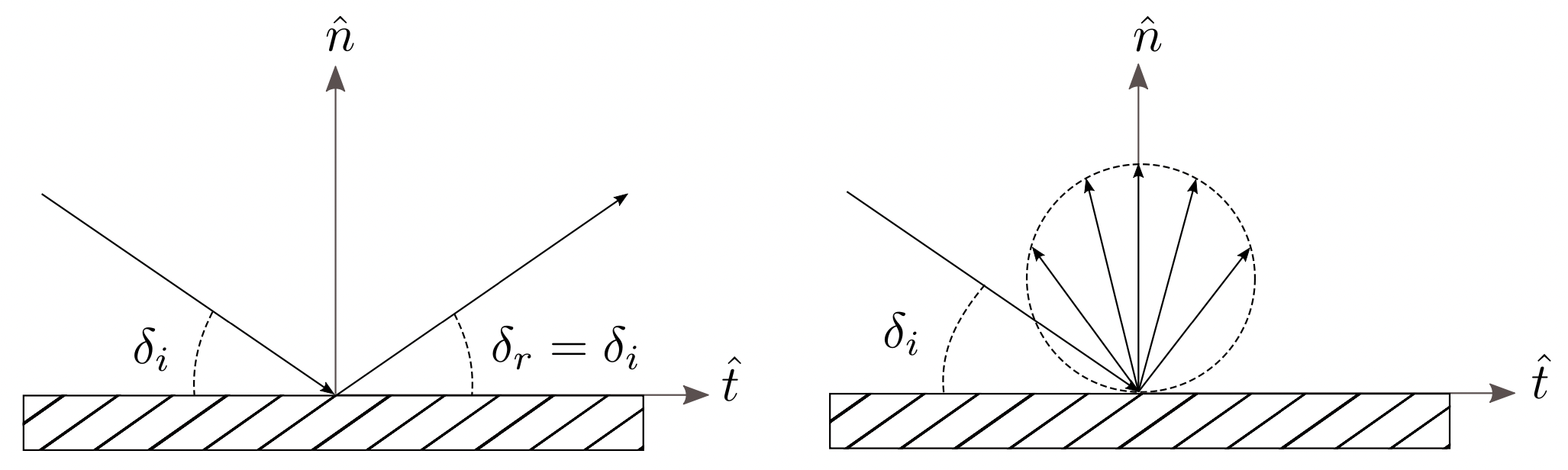}
 \caption{Mechanism of specular and diffuse re-emission~\cite{LIVIA}}
 \label{fig:SD-ref}
\end{figure}

Upon particle impact with a surface an exchange of momentum and energy takes place, and the velocity of the incoming molecule is altered by the specific local properties of the impacted surface. The Maxwell model is a simplified approach that uses two types of surface-particle interactions: specular and diffuse reflection. The schematics are shown in Fig.~\ref{fig:SD-ref} in which the velocity vector of the particles $v$ is represented by the black arrows with a thinner tip, $\hat{n}$ and $\hat{t}$ are the normal and tangential unit vectors, and $\delta_i$ and $\delta_r$ the incident and re-emission angles. 

In specular reflections, particles rebound when hitting the surface without energy exchange, the angle of reflection equals the incident angle $\delta_r = \delta_i$, and the tangential component of particle velocity $\bold{v}\hat{t}=v _t$ remains constant. Only the normal velocity component $\bold{v}\hat{n}=v_n$ is affected undergoing a complete inversion $v_t \rightarrow - v'_t$. In diffuse reflection, instead, the velocity of reflected molecules $\bold{v}'$ is independent of the incident one. Particles reach thermal equilibrium with the surface and are reflected according to a Rayleigh distribution corresponding to the wall temperature $T_{wall}$. The Maxwell model defines the accommodation coefficient $\sigma_B$, to represent the fraction between diffuse and specular reflection, and it is highly dependent on the surface material properties. The accommodation coefficient can be implemented in both the momentum and energy equations~\cite{FMF_STORCH2002} and is defined in Eq.~\ref{Eq:sigmaB}, where the energy of the incident particle is $E$, and $E'$ the energy of the reflected particle, while $E_W$ is the energy of the particle which it would have at a full accommodation to $T_{wall}$.	
\begin{equation}
 \sigma_B = \frac{E-E'}{E - E_W}
 \label{Eq:sigmaB}
\end{equation}

For realistic free molecular flows (FMF) and most surface materials, a full accommodation $\sigma_B=1$, corresponding to diffuse reflection, provides good simulation results on surface interactions ~\cite[Ch.~5,~pp.~118]{bird1994molecular}. Instead, specular reflection is for $\sigma_B=0$. 

\section{Atmospheric Model}
\begin{figure}[H]
 \centering
 \includegraphics[width=.95\textwidth, trim={2.5cm 0cm 0cm 1cm},clip]{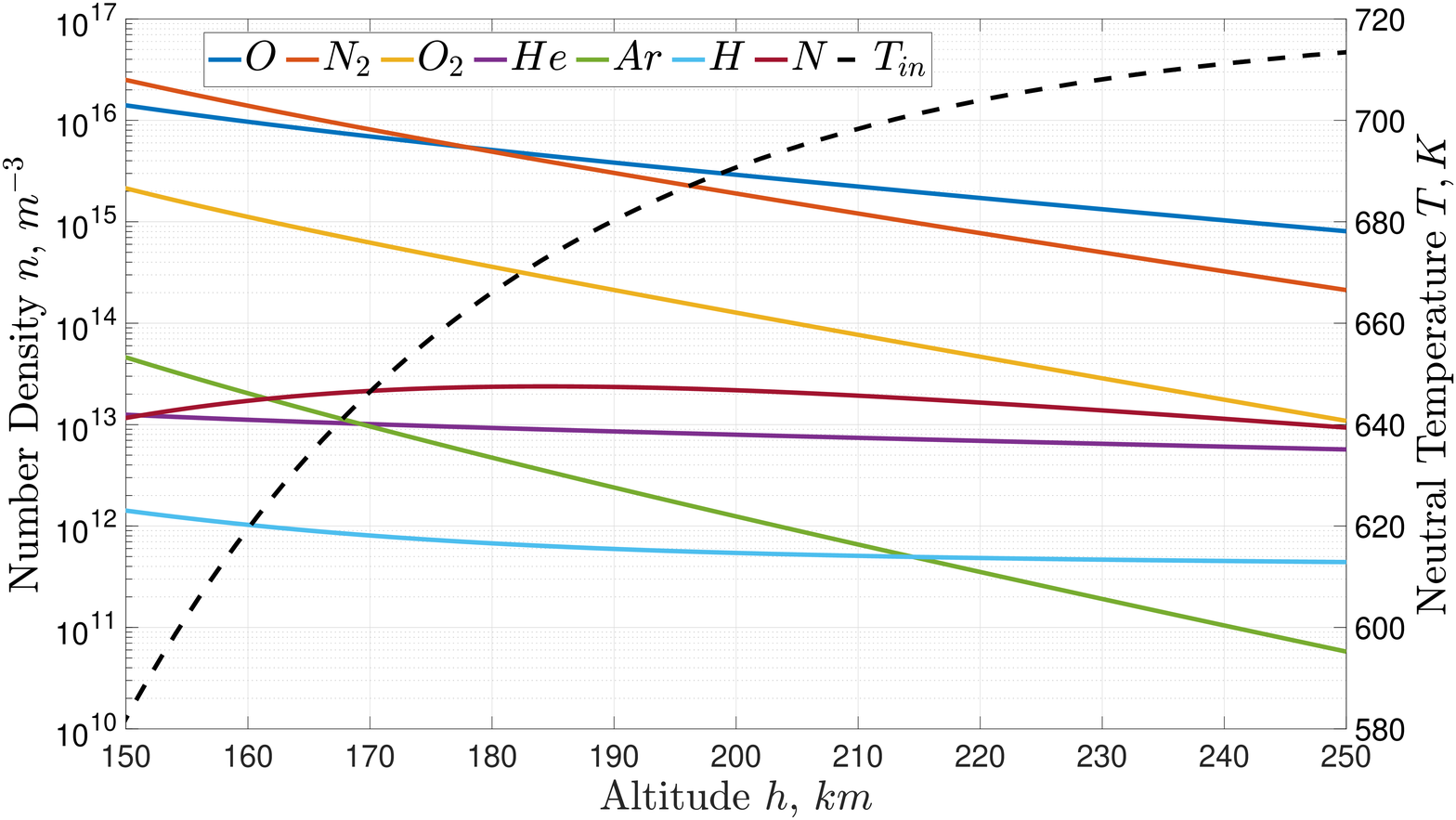}
 \caption{Atmospheric Composition in VLEO, NRLMSISE-00 Atmospheric Model: 15/02/2020 at 00:00:00, $F10.7 = 69.5$, $Ap = 4.1$, averaged $0-\SI{360}{\degree}$}
 \label{fig:nrlmsise}
\end{figure}

Among the several atmospheric models available, the NRLMSISE-00 is used as it accurately accounts for the lower thermosphere atmospheric conditions, and has been previously used for work on ABEP systems~\cite{VLEO_LEOMANNI2017,6945885,romanoacta}. In Fig.~\ref{fig:nrlmsise}, the averaged particle number densities and neutral temperature $T_{in}$ are plotted in VLEO for an equatorial orbit. The latitude is fixed to \SI{0}{\degree} while longitudes are averaged between $0-\SI{360}{\degree}$ in \SI{45}{\degree} steps. The data is extracted from the NASA Community Coordinated Modelling Center~\cite{NASA_nrlmsise00} for the 15/02/2020 at 00:00:00 corresponding to $F10.7 = 69.5$, the solar radio flux at a wavelength of $\lambda=\SI{10.7}{\centi\meter}$, and the geomagnetic index $Ap=4.1$~\label{sec:NRLMSISE}. The date is selected as it offers the latest updated solar activity data for NASA Community Coordinated Modeling Center at the beginning of the current research, corresponding to low solar activity. However, once the mission lifetime and launch date are set, the respective solar activity must be taken into account as can lead to large variations in the atmospheric environment~\cite{romanoacta}. The selected altitudes range is between $h=150-\SI{250}{\kilo\meter}$. The lower limit is due to aerodynamic drag, that starts increasing exponentially leading to very high power requirements~\cite{6945885} and also heating of the SC~\cite{romanoacta}. The upper limit is set due to the limited propellant collection, falling below thruster requirements by a typical sized SC, and where conventional electric propulsion becomes competitive against an ABEP system~\cite{di2007ram}. The dominant species in the selected altitude range are \ce{N2} and atomic oxygen (AO).

\section{The PICLas Numerical Tool}
PICLas is a numerical tool for simulating non-equilibrium gas and plasma flows developed by the Institute of Space Systems (IRS) and the Institute of Aerodynamics and Gasdynamics (IAG) at the University of Stuttgart~\cite{PICLAS}. The code is a three-dimensional, parallelized simulation framework for the coupling of, among others, DSMC and Particle-in-Cell~\cite{Munz2014662,PICLAS}. For the scope of this project, the DSMC module with the Maxwell model for surface interactions is used. Within the code, the proportion of diffuse and specular reflections is regulated by the momentum accommodation coefficient (MomentumACC). For the case MomentumACC$>R_{PICLas}$ a diffuse reflection is performed, otherwise a specular one is performed during a GSI ($R_{PICLas}$:~a given number within $R_{PICLas}=[0,1)$).
  
\subsection{Simulation Settings}
The simulation domain of the intake is discretized using unstructured hexahedral cells. Due to the symmetry of the intake design, see Fig.~\ref{fig:Gen_Diag}, and for an incoming flow parallel to $z$, symmetry planes normal to $x$ and $y$ are used to reduce the mesh size and therefore optimize the simulation efficiency. Additionally, more optimized concepts are explored with two different mesh setups and simulation times to verify the results~\cite{EspinosaOrozco1543422}. The ABEP VLEO altitude range of $h=150-\SI{250}{\kilo\meter}$ corresponds to the orbital velocity $v_{SC}\sim\SI{7.8}{\kilo\meter\per{\second}}$~\cite{romanoacta}.
The free stream conditions at VLEO are in the free molecular flow regime~\cite{tilman,romanoacta}, and the particle flow can be treated as hyperthermal for the selected $AR$ in this study~\cite{Intake_YANWU2015,tilman}. This is valid at low temperatures when the random thermal motion of the incoming particles is negligible compared to the SC velocity $v_{SC}$. Therefore, the incoming flow is collimated entering the intake with a single free stream bulk velocity $v_{in}(h)=v_{SC}(h)$ at a specific incoming angle $\alpha$. Assuming a circular orbit, $v_{SC}(h)$ is calculated as in Eq.~\ref{eq:v_orb}, where $\mu_{E} = \SI{3.986E+15}{\kilo\meter\per{\second^2}}$ is the Earth's gravitational parameter, and its average radius $R_{E} = \SI{6371}{\kilo\meter}$.
\begin{equation}
    v_{SC}(h) = \sqrt{\frac{\mu_{E}} {R_{E} + h}}
    \label{eq:v_orb}
\end{equation}
Finally, the temperature of the intake walls $T_{wall}$ is set to that of a SC in LEO $T_{wall}=\SI{300}{\kelvin}$~\cite{romanoiepc,romanoacta}. Investigation based on variation of $T_{wall}$ in the intake has been performed previously, but have shown no significant influence on intake performance~\cite{tilman}. The flow parameters used as input in PICLas are shown in Tab.~\ref{tab:DSCM__flow}, and are extracted from the NRLMSISE-00 atmospheric model as specified in Sec.~\ref{sec:NRLMSISE}. Finally, all the species shown in Fig.~\ref{fig:nrlmsise}, are accounted in the simulations with their respective $n_i$.

\begin{table}[H]
\centering
\caption{PICLas Simulations Flow Parameters based on NRLMSISE-00 Atmospheric Model}
~\\
\label{tab:DSCM__flow}
\begin{tabular}{ p{0.1\textwidth}<{\raggedright} p{0.12\textwidth}<{\centering}  p{0.12\textwidth}<{\centering}  p{0.12\textwidth}<{\centering} p{0.18\textwidth}<{\centering}  }
\toprule
\textbf{$h$} & \textbf{$T_{in}$} & \textbf{$T_{wall}$} &  \textbf{$v_{SC}$} & \textbf{$n_{in,total}$}   \\ 
\SI{}{\kilo\meter} & \SI{}{\kelvin} & \SI{}{\kelvin} & \SI{}{\meter\per{\second}} & $\SI{}{1/\meter^3}$ \\ 
\midrule
150 & 582.0 & 300 & 7818.2 & 4.131E+16 \\
180 & 666.1 & $"$ & 7800.3 & 1.042E+16 \\
200 & 690.7 & $"$ & 7788.5 & 4.967E+15\\
220 & 703.9 & $"$ & 7776.6 & 2.560E+15\\
250 & 713.4 & $"$ & 7759.0 & 1.045E+15\\
\bottomrule
\end{tabular}
\end{table}

The intake concepts based on diffuse reflecting materials are represented by MomentumACC$=1.0$, while those based on specular reflecting materials by MomentumACC$=0.0$. In the hybrid concepts, each surface boundary condition (B.C.) is singularly defined as either diffuse or specular. As a final remark, all the intakes are based on an $A_{out}$ that is defined by the selected reference thruster~\cite{romanoacta}.
\label{sec:momacc}

\section{Intake Based on Diffuse Reflecting Materials}
\label{sec:diff}
\begin{figure}[H]
    \centering
    \includegraphics[width=.9\textwidth, trim={4cm 1.5cm 8cm 3cm},clip]{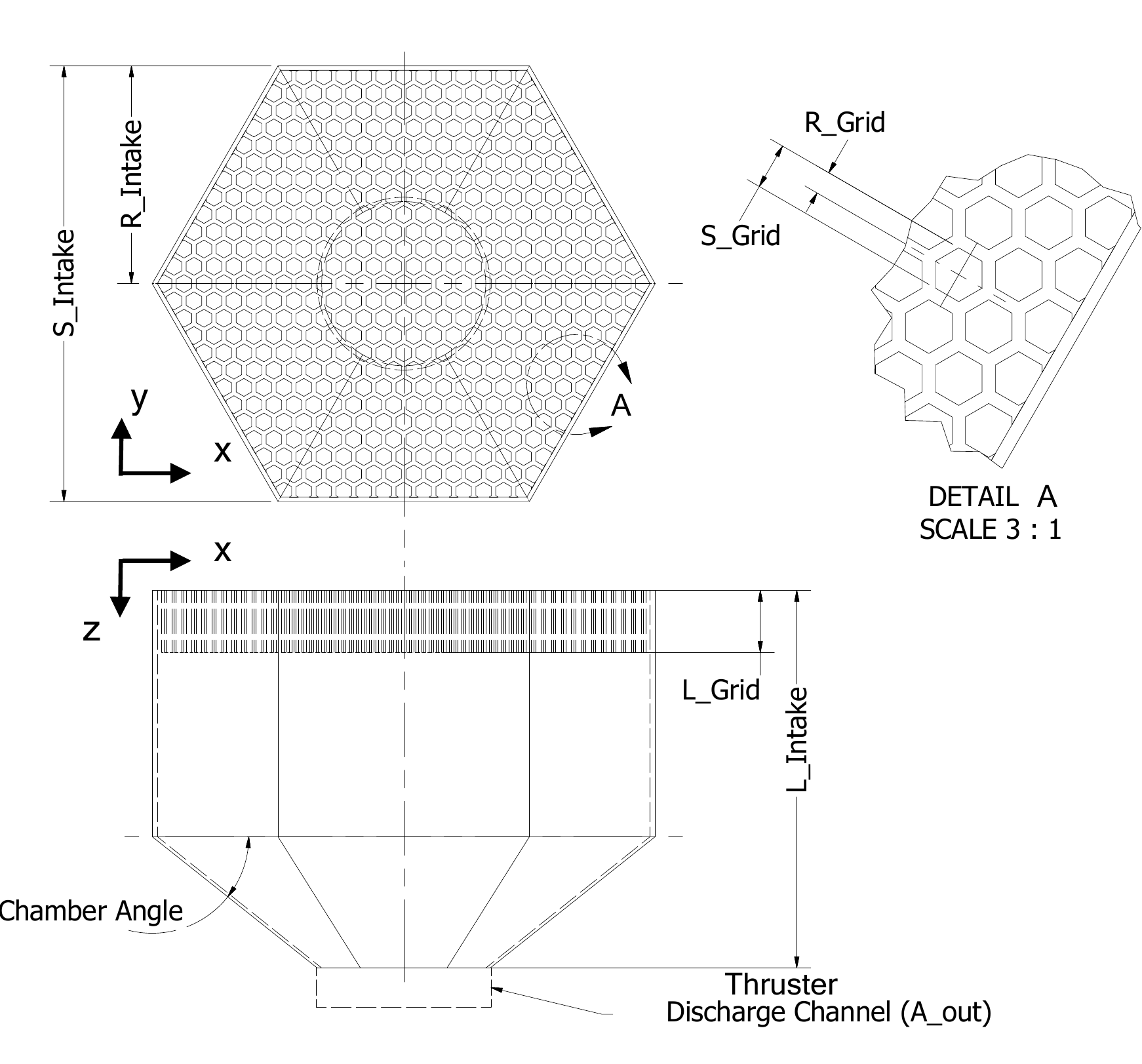}
    \caption{Diffuse Intake Design Schematics}
    \label{hex_ar}
\end{figure}

The previous work performed at IRS on diffuse reflection-based intake designs concluded that hexagonal geometries are the most adequate for the intake front grid geometry due to the lower backflow transmission probabilities for single and multiple ducts~\cite{tilman}. Such analysis included an analytical model, the balance model (BM), based on transmission probabilities~\cite{romanoiepc}, and simulations for optimal aspect ratios ($AR$) of the duct grid geometry, defined as the ratio between duct length and square root of its cross section. Finally, the results have been verified via simulations performed with PICLas~\cite{romanoiepc,romanosp2016,tilman}. Based on the previous work, different diffuse reflection-based intake concepts are hereby analysed by implementing hexagonal grid geometries for 3D investigation and tailored, a given $A_{out}$, to the RF Helicon-based plasma thruster as reference~\cite{romanoacta2}. Various grid $AR$ are simulated based on diffuse reflecting properties to evaluate the effect on $\eta_c$. Finally, the hexagonal outer geometry is also implemented to maximize the filling ratio (FR) of the duct structure as $A_{in}=A_{Intake} FR$, and to facilitate flow simulations. The baseline intake design used for the investigation is shown within Fig.~\ref{hex_ar}.

\subsection{Design Improvement Settings and Procedure}
The design improvement process is iterative. The starting case is set for an altitude $h=\SI{150}{km}$ at an incoming angle $\alpha=\SI{0}{\degree}$. Based on this different (intake) cases are designed, simulated, and evaluated for better understating the scattering dynamics inside the hexagonal ducts in terms of $\eta_c$. The first iteration affects the variation of overall intake size and chamber angles, with a fixed duct $AR$ based on previous studies~\cite{romanoiepc,romanosp2016,tilman}. Then, the designs providing $\eta_c \ge 0.3$ are further adapted to the hexagonal outer shape. In the second iteration step, different duct $AR$ are simulated, while keeping chamber angle, intake length, and intake diameter constant. Finally, the concepts with the highest $\eta_c$ are selected for further investigation and the respective collectible $\dot{m}_{thr}$ and pressure distribution of $p_{ch}$ are taken into account. 

\subsection{Simulation Results}
The three selected designs are named Diffuse-Nominal (D-N), Diffuse-Multiple Duct AR (D-MDAR), and Diffuse-Maximum Filling Ratio (D-MFR). The respective performance and settings are shown in Tab.~\ref{tab:results-hexagonal}. The D-N case is derived directly from the enhanced funnel design (EFD)~\cite{romanoiepc,romanosp2016,tilman}, adapted to an hexagonal shape, and hereby optimized for the RF Helicon-based plasma thruster. This case is the baseline for all the upcoming concepts. 

\begin{figure}[H]
    \centering
    \includegraphics[width=.9\textwidth, trim={0cm 4.5cm 0.5cm 3cm},clip]{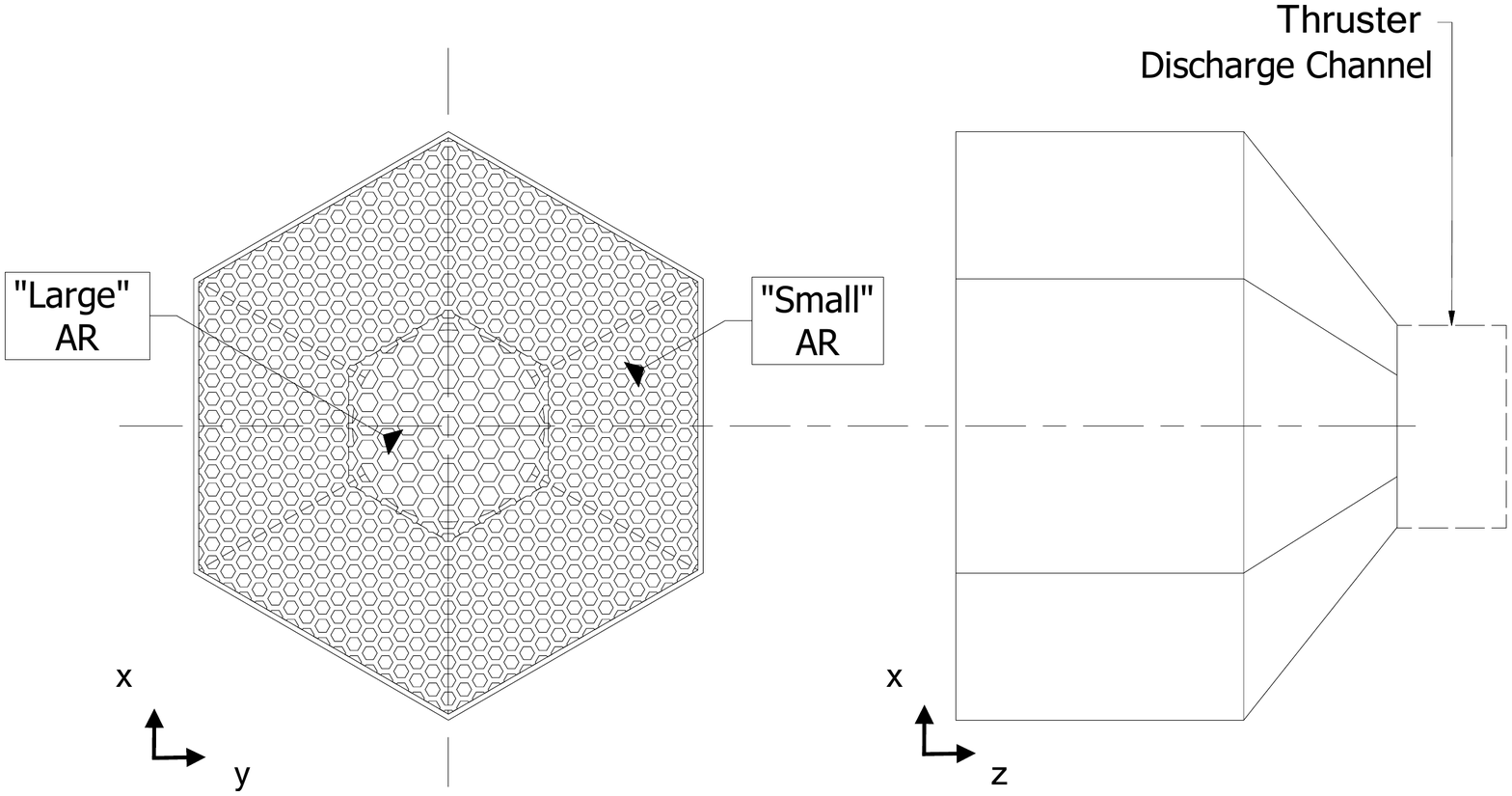}
    \caption{Frontal and Lateral Views of the D-MDAR Intake Design}
    \label{MAR}
\end{figure}

The D-MDAR design features multiple duct $AR$ to increase the effective intake area $A_{in}$ directly in front of the thruster's discharge channel with larger $AR$ ducts, while smaller ducts are at the outer perimeter, as this region is expected to have larger backflow. The schematics displaying the frontal view of the D-MDAR is shown in Fig.~\ref{MAR}. 

The D-MFR case is designed to maximize $A_{in}$, while still taking advantage of the backflow mitigation of the duct section. The number of ducts on the intake is minimized and their $AR$ equal, finally reducing the cross-sectional area of the duct geometry while keeping the same wall thickness.

\begingroup
\setlength{\tabcolsep}{6pt} 
\renewcommand{\arraystretch}{1.3} 
\begin{table}[H]
\centering
\caption{Diffuse Reflection-Based Cases Results,~$h=\SI{150}{\kilo\meter}$}
\label{tab:results-hexagonal}
\resizebox{\textwidth}{!}{\begin{tabular}{lccccccc}
\toprule
 &  $A_{Intake}$ & FR & $A_{in}$ & $\dot{N}_{in}$ & $\dot{N}_{out}$ & $\eta_{c}$ & $\dot{m}_{thr}$ \\ 
 & $\SI{}{\meter^2}$ & $\%$ & $\SI{}{\meter^2}$ & $\SI{}{1/\second}$ & $\SI{}{1/\second}$ & - & $\SI{}{\milli\gram\per{\second}}$\\ \midrule
D-N & 0.008 & 50.6 & 0.004 & 1.31E+18 & 4.15E+17 & 0.317 & 0.0193 \\ 
D-MDAR & 0.008 & 51.8 & 0.004 & 1.34E+18 & 6.14E+17 & 0.458 & 0.0240 \\ 
D-MFR & 0.004 & 92.6 & 0.004 & 1.24E+18 & 5.43E+17 & 0.437 & 0.0212 \\ \bottomrule
\end{tabular}}
\end{table}
\endgroup

\begin{figure}[H]
    \centering
    \includegraphics[width=.85\textwidth]{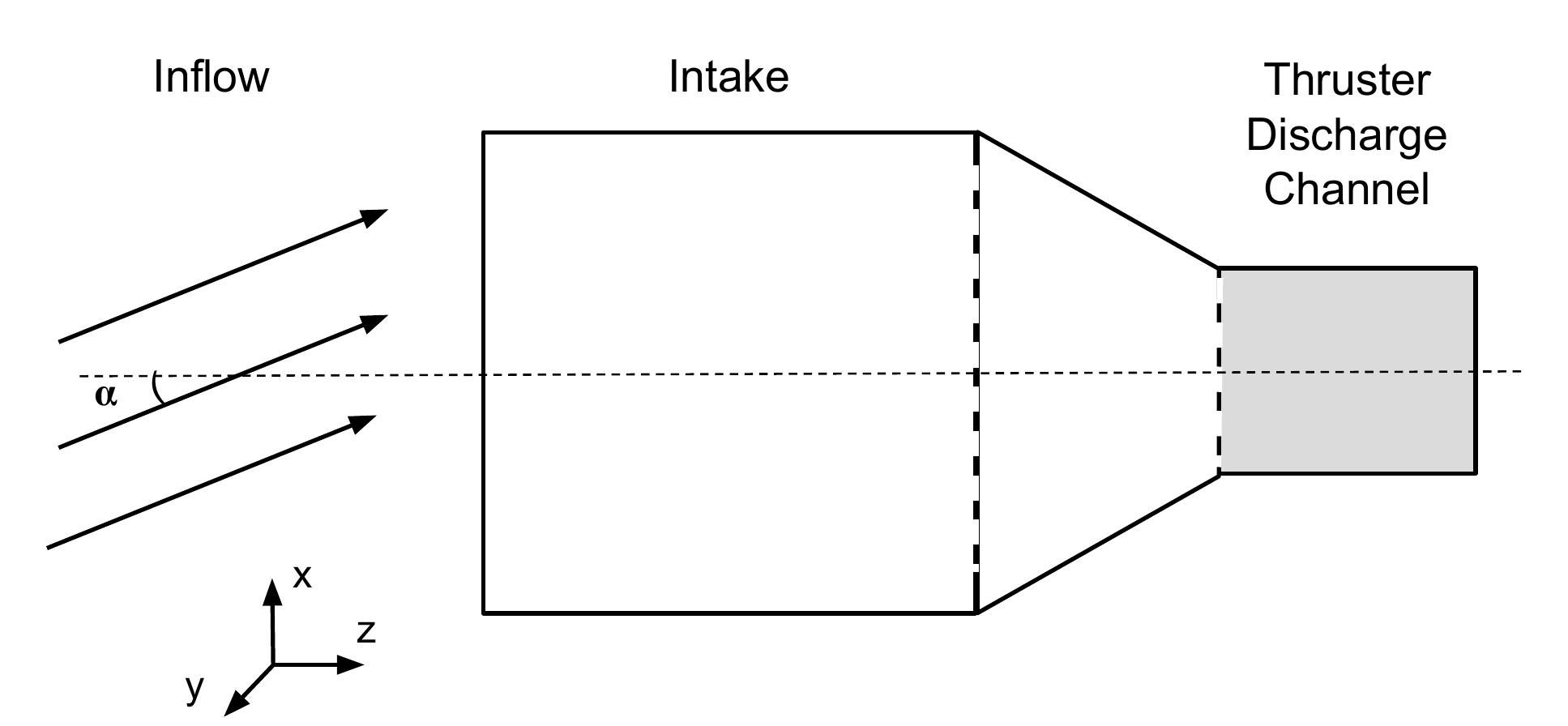}
    \caption{Incoming Flow Angle $\alpha$ Diagram.}
    \label{FIG:ALPHA}
\end{figure}

As listed in Tab.~\ref{tab:results-hexagonal}, the D-MDAR case provides the highest performance in terms of both $\dot{m}_{thr}$ and $\eta_c$. Although, the D-MFR also provides $\eta_c > 0.4$ with only half frontal area $A_{Intake}$, this intake design provides a $10\%$ lower $\dot{m}_{thr}$ when compared with D-MDAR case. 
Different incoming flow angles $\alpha$, see Fig.~\ref{FIG:ALPHA}, are simulated to assess D-MDAR performance variations at conditions in which the SC may not be able to maintain flow-oriented attitude, or due to the presence of strong not-aligned atmospheric winds as seen by GOCE~\cite{GOCECD}.

In Tab.~\ref{tab:c6-miss}, the results of this analysis are presented, displaying the relative $\Delta \eta_c$ variation of $\eta_c$ vs $\alpha$.

\begingroup
\setlength{\tabcolsep}{6pt} 
\renewcommand{\arraystretch}{1.2} 
\begin{table}[H]
\centering
\caption{D-MDAR Case, Flow Misalignment Analysis,~$h=\SI{150}{\kilo\meter}$}
\label{tab:c6-miss}
\begin{tabular}{lccccc}
\toprule
Geometry &  $\alpha$& $\dot{N}_{out}$ & $\eta_{c}$ & $\dot{m}_{thr}$ & $\Delta \eta_c$\\
&  $\SI{}{\degree}$ & $\SI{}{1/\second}$ & - & $\SI{}{\milli\gram\per{\second}}$ & $\%$\\
 \toprule
\multirow{5}{*}{D-MDAR} & 0 & 6.11E+17  & 0.456 & 0.0240  & - \\ 
 & 5 & 5.07E+17 & 0.378  & 0.0198 & $-17$\\ 
 & 10 & 3.62E+17 & 0.270 & 0.0140  &$-41$\\ 
 & 15 & 2.60E+17 & 0.194 & 0.0099  &$-58$\\ 
 & 20 & 2.01E+17 & 0.150 & 0.0077  &$-67$\\ 
 \bottomrule
\end{tabular}
\end{table}
\endgroup

\section{Intake Based on Specular Reflecting Materials}
\label{sec:spec}
\begin{figure}[H]
    \centering
    \includegraphics[width=.85\textwidth, trim={4cm 3.5cm 2cm 5cm},clip]{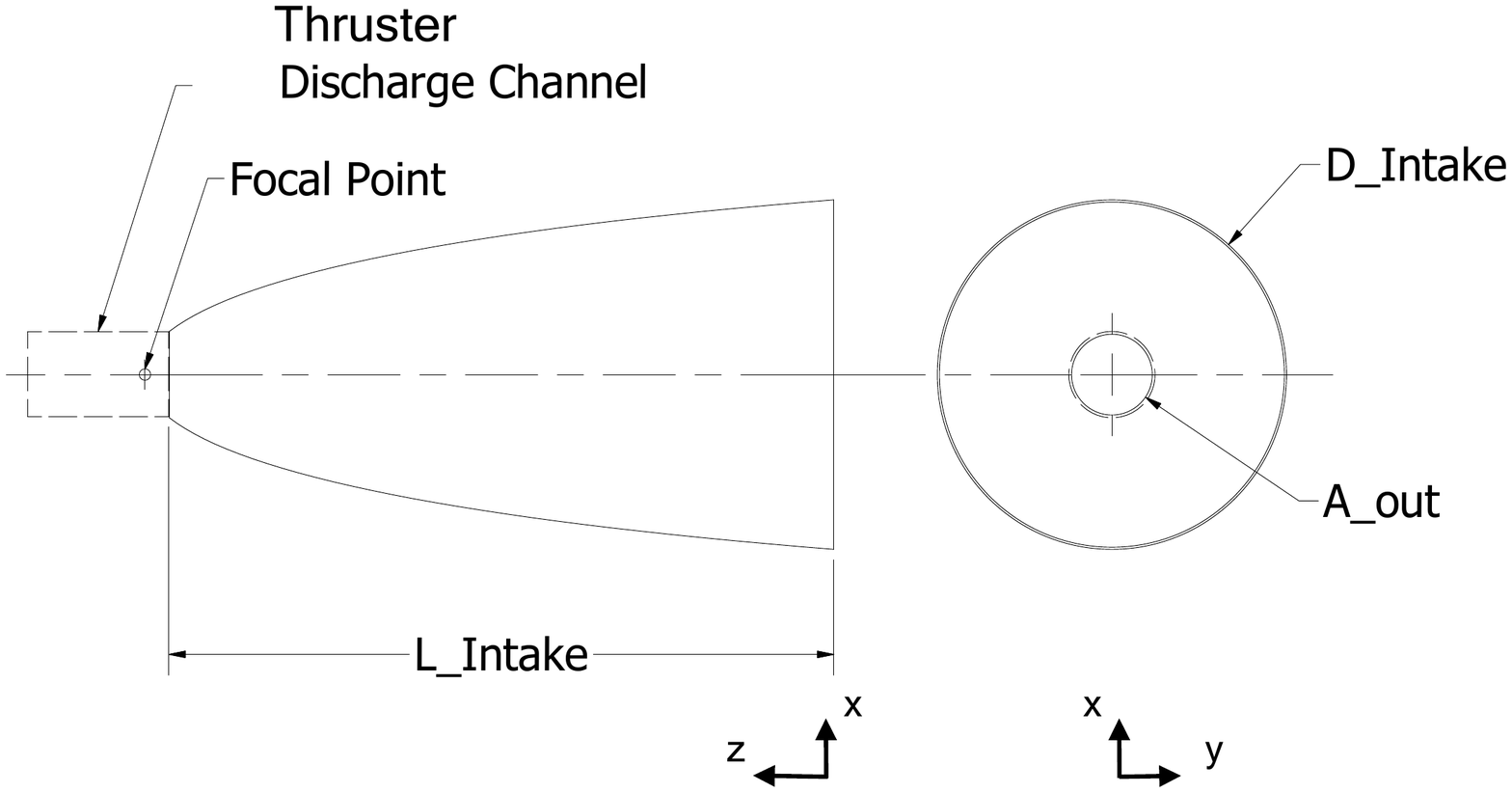}
    \caption{Specular Intake Concept.}
    \label{par_ar}
\end{figure}

The intake designs based on specular reflecting materials, $\sigma_B = 0$, are aimed to maximize $\eta_c$ by taking advantage of specular scattering surfaces in combination with optical-based parabolic geometries. The focal point of the parabola is designed to direct the incoming particles to the thruster's discharge channel. A general schematic of the concept is shown in Fig.~\ref{par_ar}.
 
\begin{figure}[H]
    \centering
    \includegraphics[width=0.85\textwidth, trim={2.5cm 3cm 4cm 3cm},clip]{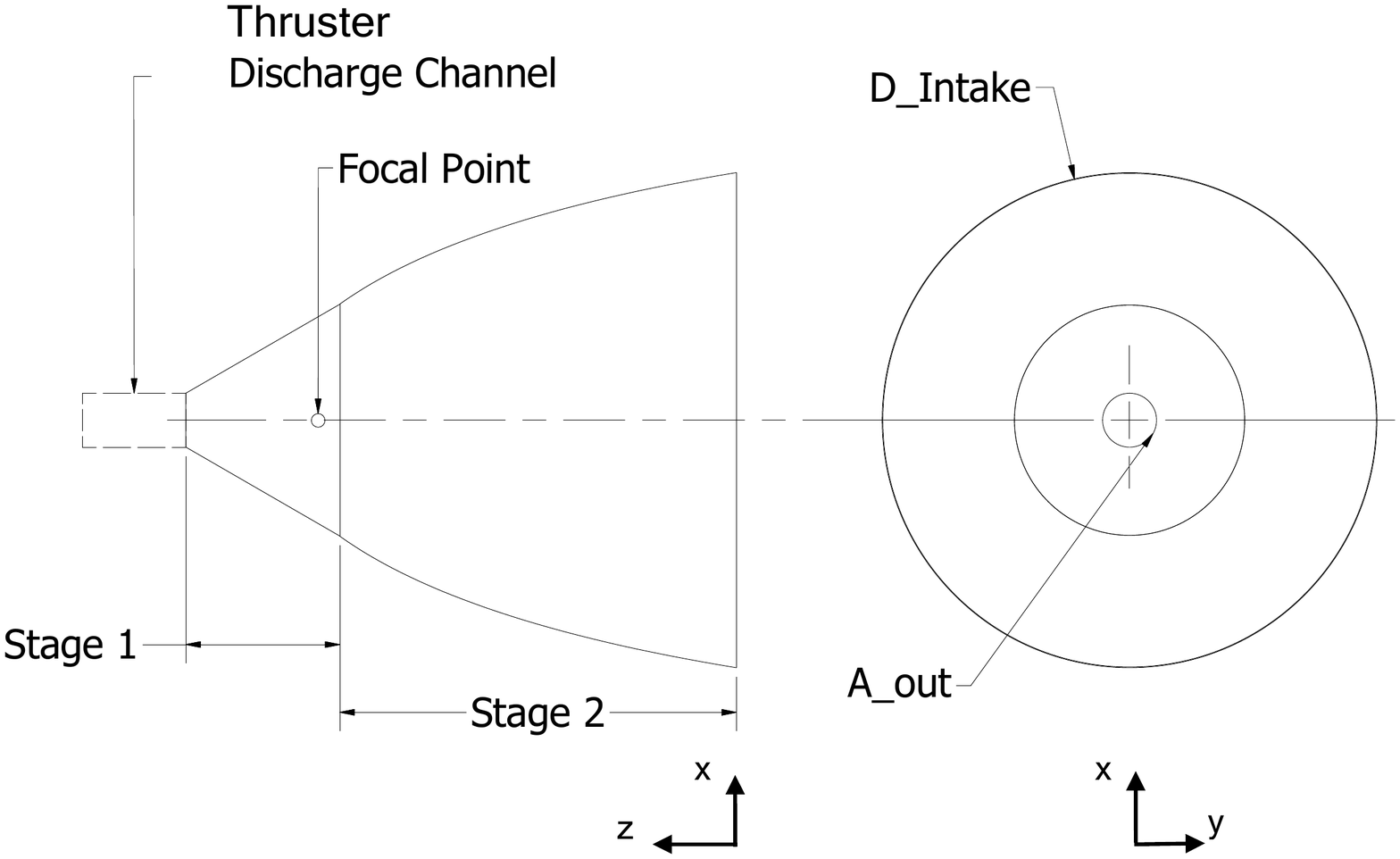}
    \caption{S-MS Intake Concept.}
    \label{ext_par}
\end{figure}

Furthermore, concepts with an additional stages are developed, named Specular Multiple-Stage intakes (S-MS), to combine different area ratios to possibly increase $\dot{m}_{thr}$, see Fig.~\ref{ext_par}.

\subsection{Design Improvement Settings and Procedure}
The design improvement process is iterative. The base case is set for $h=\SI{150}{\kilo\meter}$ at an incoming angle $\alpha=\SI{0}{\degree}$. Each case is tailored to the given $A_{out}$ of the reference thruster~\cite{romanoacta2}, and has different geometric parameters which affect the focal point location of the parabola targeting the maximization of $\eta_c$. Furthermore, the cases providing the highest $\eta_c$ are the investigated aiming to maximizing $\dot{m}_{thr}$ and $p_{ch}$. Concerning the S-MS intake concept, different geometrical stage combinations are simulated, based on the improved purely specular designs, to maximize $\dot{m}_{thr}$ and $p_{ch}$ while maintaining $\eta_c > 0.3$.

\subsection{Simulation Results}
\begin{figure}[H]
    \centering
    \includegraphics[width=.85\textwidth, trim={1cm 1cm 1.5cm 0cm},clip]{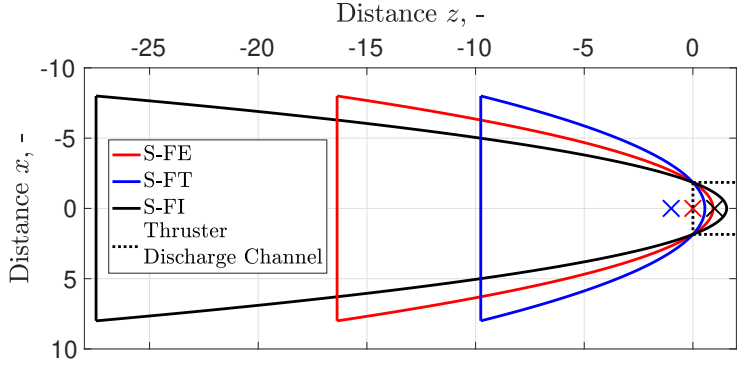}
    \caption{Parabolic Intake Configurations (normalized units).}
    \label{fig:parabolic_focus}
\end{figure}

The highest performance cases for the intake based on specular reflecting materials are shown in Tab.~\ref{tab:results-parabolic}. The naming of the cases refers to the focal point position of the parabola in respect to the thruster discharge channel, the schematics is shown in Fig.~\ref{fig:parabolic_focus}. 
\begin{itemize}
\item S-FE: focus at the boundary between the intake and discharge channel;
\item S-FT: focus located inside the discharge channel of the thruster;
\item S-FI: focus within the intake itself.
\end{itemize}

Finally, the S-FT-D case includes a frontal duct section to take advantage of both the parabolic optical properties, and the low backflow transmission probabilities of the ducts structure. The parabolic geometry is shared with the S-FT case, but, due to the duct section, the $A_{in}$ of the S-FT-D case is smaller.

\begingroup
\setlength{\tabcolsep}{6pt} 
\renewcommand{\arraystretch}{1.3} 
\begin{table}[H]
\centering
\caption{Specular Cases Results $h=\SI{150}{\kilo\meter}$.}
\label{tab:results-parabolic}
\resizebox{\textwidth}{!}{\begin{tabular}{lccccccc}
\toprule
 &  $A_{Intake}$ & FR & $A_{in}$ & $\dot{N}_{in}$ & $\dot{N}_{out}$ & $\eta_{c}$ & $\dot{m}_{thr}$ \\ 
 & $\SI{}{\meter^2}$ & $\%$ & $\SI{}{\meter^2}$ & $\SI{}{1/\second}$ & $\SI{}{1/\second}$ & - & $\SI{}{\milli\gram\per{\second}}$\\
 \midrule
S-FT & 0.019 & 100.0 & 0.019& 6.33E+18 & 5.97E+18 & 0.943 & 0.232 \\
S-FT-D & 0.019 & 88.5 & 0.017 & 5.61E+18 & 4.97E+18 & 0.887 & 0.193 \\
S-FI & 0.041 & 100.0 & 0.041 & 1.33E+19 & 7.83E+18 & 0.591 & 0.309 \\
S-FE & 0.025 & 100.0 & 0.025 & 8.11E+18 & 7.15E+18 & 0.882 & 0.279 \\
\bottomrule
\end{tabular}}
\end{table}
\endgroup

Furthermore, the results for the highest performance cases of the S-MS concept is shown in Tab.~\ref{tab:results-parabolic2}. The S-MS-F and S-MS-B cases are both applied to the S-FT design described above. In the S-MS-F, $A_{Intake}\sim A_{in}$ is increased by adding a second stage at the front of the S-FT case parabola, while in the S-MS-B case, the second stage is added at the back of the S-FT parabola.

\begingroup
\setlength{\tabcolsep}{6pt} 
\renewcommand{\arraystretch}{1.3} 
\begin{table}[H]
\centering
\caption{S-MS Cases Results $h=\SI{150}{\kilo\meter}$.}
\label{tab:results-parabolic2}
\resizebox{\textwidth}{!}{\begin{tabular}{lccccccc}
\toprule
 &  $A_{Intake}$ & FR & $A_{in}$ & $\dot{N}_{in}$ & $\dot{N}_{out}$ & $\eta_{c}$ & $\dot{m}_{thr}$ \\ 
 & $\SI{}{\meter^2}$ & $\%$ & $\SI{}{\meter^2}$ & $\SI{}{1/\second}$ & $\SI{}{1/\second}$ & - & $\SI{}{\milli\gram\per{\second}}$\\
\midrule
S-MS-F & 0.090 & 100.0 & 0.090 & 2.91E+19 & 5.90E+18 & 0.203 & 0.229 \\
S-MS-B & 0.090 & 100.0 & 0.090 & 2.91E+19 & 9.71E+18 & 0.334 & 0.384 \\
\bottomrule
\end{tabular}}
\end{table}
\endgroup

\begin{figure}[H]
    \centering
    \includegraphics[width=.8\textwidth]{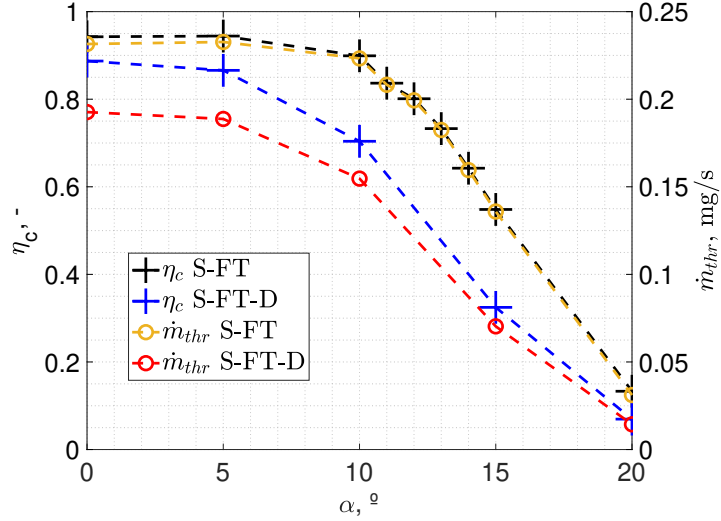}
    \caption{S-FT and S-FT-D Cases $\eta_c$ and $\dot{m}_{thr}$ vs. $\alpha$.}
    \label{fig:Ralt_pGrid}
\end{figure}

The designs providing the highest $\eta_c$, the S-FT and the S-FT-D, are further investigated to evaluate the performance sensitivity to flow misalignment $\alpha$.
Based on the results of the diffuse duct simulations, it is expected that the duct geometry could prove useful on mitigating the effects of $\alpha>\SI{0}{\degree}$, as particles could be captured by the duct geometry and redirected to the parabolic section. The collection efficiency $\eta_c$  and the output mass flow $\dot{m}_{thr}$ at different $\alpha$ are shown in Fig.~\ref{fig:Ralt_pGrid}, where the S-FT case is examined in a closer manner between $\SI{10}{\degree} > \alpha > \SI{15}{\degree}$ to better understand the influence of $\alpha$ to the parabolic scattering dynamics.

\section{Hybrid Intake Based on Specular and Diffuse Reflecting Materials}

\label{sec:hybrid}
\begin{figure}[H]
    \centering
    \includegraphics[scale = 0.4, trim={1.5cm 3cm 2cm 3.5cm},clip]{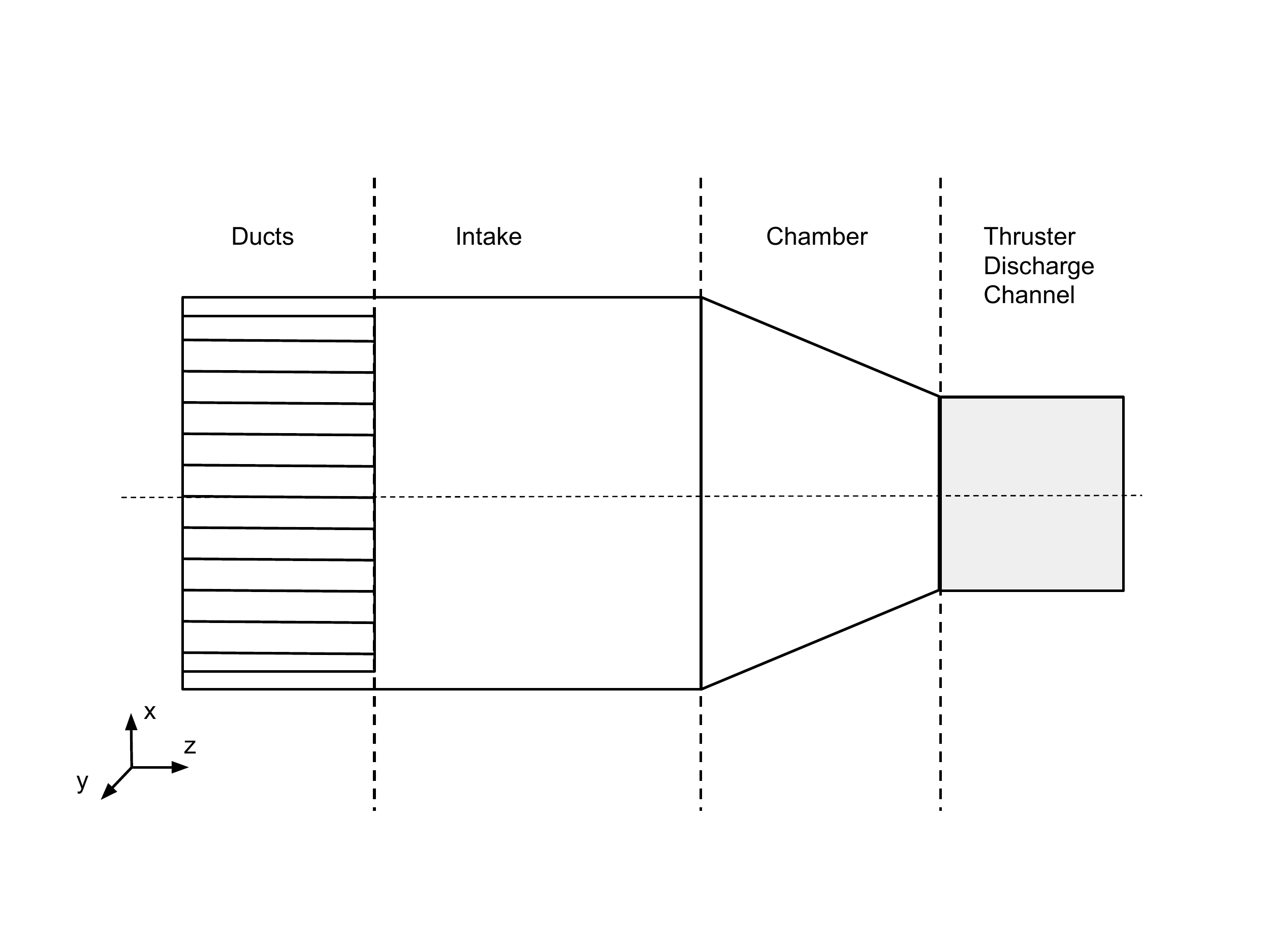}
    \caption{Intake Boundary Condition Sections Scheme}
    \label{intake_SQ}
\end{figure}

The hybrid intake concepts are based on a multiple stage design and foresee a (hybrid) combination of diffuse $\sigma_B = 1$ and specular $\sigma_B = 0$ reflecting materials. The schematics showing the different intake sections at which the B.C. are applied is shown in Fig.~\ref{intake_SQ}.

\subsection{Design Improvement Settings and Procedure}
The hybrid intake concepts are based on the already improved designs presented in Sec.~\ref{sec:diff}, and Sec.~\ref{sec:spec}. Therefore, the design improvement process is focused only on the overall intake performance effect due to different combinations of GSI properties.

At first, the geometry represented by the S-FT case is simulated with all of its surfaces that can have both diffuse and specular scattering. This is achieved by varying $\sigma_B$, therefore by changing the MomentumACC in the PICLas parameter file. %
Finally, it is assumed that for MomentumACC $= 0.1$ there is a $10\%$ probability of diffuse reflection, thus, the surface is assumed to have $90\%$ specular surface properties.

\subsection{Simulation Results}
The results of the hybrid intake simulations are shown in Tab.~\ref{tab:results-hybrid}. The acronym SP indicates that the B.C. applied to the Intake or/and Chamber section is simulated with specular scattering while DI indicates diffuse scattering. The HI case refers to the hexagonal intake with a duct structure similar to the one in the baseline diffuse intake design of Fig.~\ref{hex_ar}. The S-FT-D and S-FI share the same geometric parameters of their pure specular versions, but each B.C. is simulated according to Tab.~\ref{tab:results-hybrid}.

\begingroup
\setlength{\tabcolsep}{6pt} 
\renewcommand{\arraystretch}{1.3} 
\begin{table}[H]
\centering
\caption{Hybrid Cases Results (S-FT)~$h=\SI{150}{\kilo\meter}$.}
\label{tab:results-hybrid}
\resizebox{\textwidth}{!}{\begin{tabular}{lccccccc}
\toprule
&  $A_{Intake}$ & FR & $A_{in}$ & $\dot{N}_{in}$ & $\dot{N}_{out}$ & $\eta_{c}$ & $\dot{m}_{thr}$ \\ 
& $\SI{}{\meter^2}$ & $\%$ & $\SI{}{\meter^2}$ & $\SI{}{1/\second}$ & $\SI{}{1/\second}$ & - & $\SI{}{\milli\gram\per{\second}}$\\
\midrule
\textbf{HI} & \multirow{3}{*}{0.004} & \multirow{3}{*}{87.9} & \multirow{3}{*}{0.003} & \multirow{3}{*}{1.18E+18} & \multirow{3}{*}{3.22E+17} & \multirow{3}{*}{0.272}  & \multirow{3}{*}{0.0140} \\ 
SP: Chamber &  &  &  &  &  &  &\\ 
DI: Intake, Ducts &  &  &  &  &  &  &\\ 
\midrule
\textbf{HI} & \multirow{3}{*}{0.004} & \multirow{3}{*}{87.9} & \multirow{3}{*}{0.003} & \multirow{3}{*}{1.18E+18} & \multirow{3}{*}{3.47E+17} & \multirow{3}{*}{0.294}  & \multirow{3}{*}{0.0151} \\ 
SP: Intake, Chamber &  &  &  &  &  &  &\\ 
DI: Ducts &  &  &  &  &  &  &\\ 
\midrule
\textbf{S-FT-D} & \multirow{3}{*}{0.019} & \multirow{3}{*}{88.5} & \multirow{3}{*}{0.017} & \multirow{3}{*}{5.61E+18} & \multirow{3}{*}{2.46E+18} & \multirow{3}{*}{0.439} & \multirow{3}{*}{0.098}\\ 
SP: Intake, Chamber &  &  &  &  &  &  &\\ 
DI: Ducts &  &  &  &  &  &  &\\ 
\midrule
\textbf{S-FI} & \multirow{3}{*}{0.090} & \multirow{3}{*}{100.0} & \multirow{3}{*}{0.090} & \multirow{3}{*}{2.91E+19} & \multirow{3}{*}{7.28E+18} & \multirow{3}{*}{0.250}  & \multirow{3}{*}{0.287} \\ 
SP: Intake &  &  &  &  &  &  &\\ 
DI: Chamber, Ducts &  &  &  &  &  &  &\\ 
\bottomrule
\end{tabular}}
\end{table}
\endgroup

\begin{figure}[H]
    \centering
    \includegraphics[width=\textwidth]{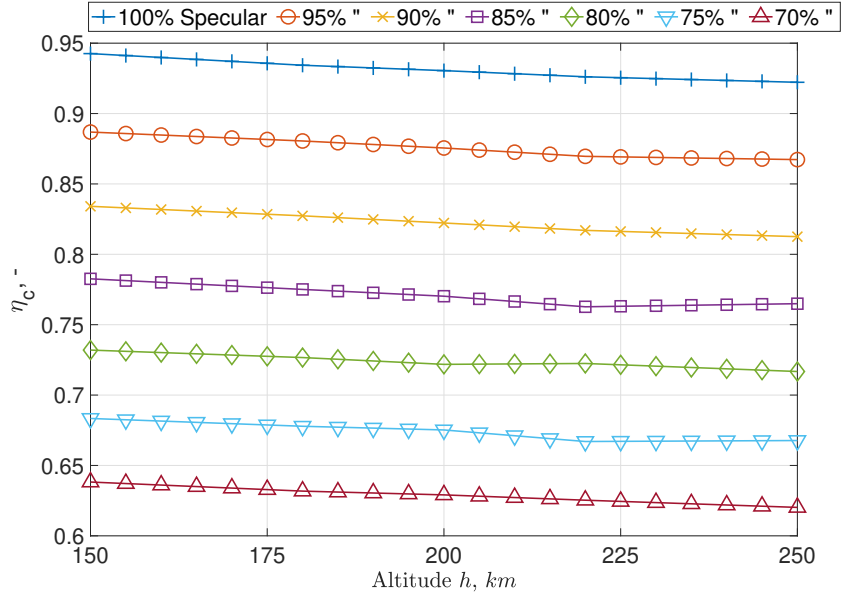}
    \caption{S-FT Intake Case, $\eta_c$ for Different Accommodation Coefficients vs. Altitude} 
    \label{fcase10_R}
\end{figure}

Additionally, the S-FT case is simulated with different level $\%$ of material specular reflection over VLEO altitudes, the respective results for $\eta_c$ are shown in Fig.~\ref{fcase10_R}. Such simulation aims to resemble the effect of different GSI properties for the parabolic intake concept providing the highest $\eta_c$. In this set of simulations, each intake section shares the same B.C. GSI property.

\section{Discussion}
This section discusses the results obtained for the different design improvements of specular, diffuse, and hybrid intake concepts.

\subsection{Diffuse Intake Improvement}
The diffuse intake improvement resulted in two designs to be the most effective for propellant collection: the D-MDAR, and the D-MFR. Both provide the highest $\eta_c=0.437-0.458$ and $\dot{m}_{thr}=0.0212-\SI{0.0240}{\milli\gram\per{\second}}$, while the second has a maximized $FR$ therefore requiring less $A_{Intake}$ which might be positive for reducing the SC frontal area, and therefore the drag. The addition of a frontal section of small ducts enhances $\eta_c$ by operating as molecular trap. The sensitivity analysis on the flow misalignment represented by the angle $\alpha$ shows that a $<50\%$ reduction of $\eta_c$ is achieved for $\alpha=\SI{10}{\degree}$, while this reduction is below $20\%$ for $\alpha<\SI{5}{\degree}$.

\subsection{Specular Intake Improvement}
\begin{figure}[H]
 \centering
 \includegraphics[width=.95\textwidth, trim={0cm 1cm 0cm 3cm},clip]{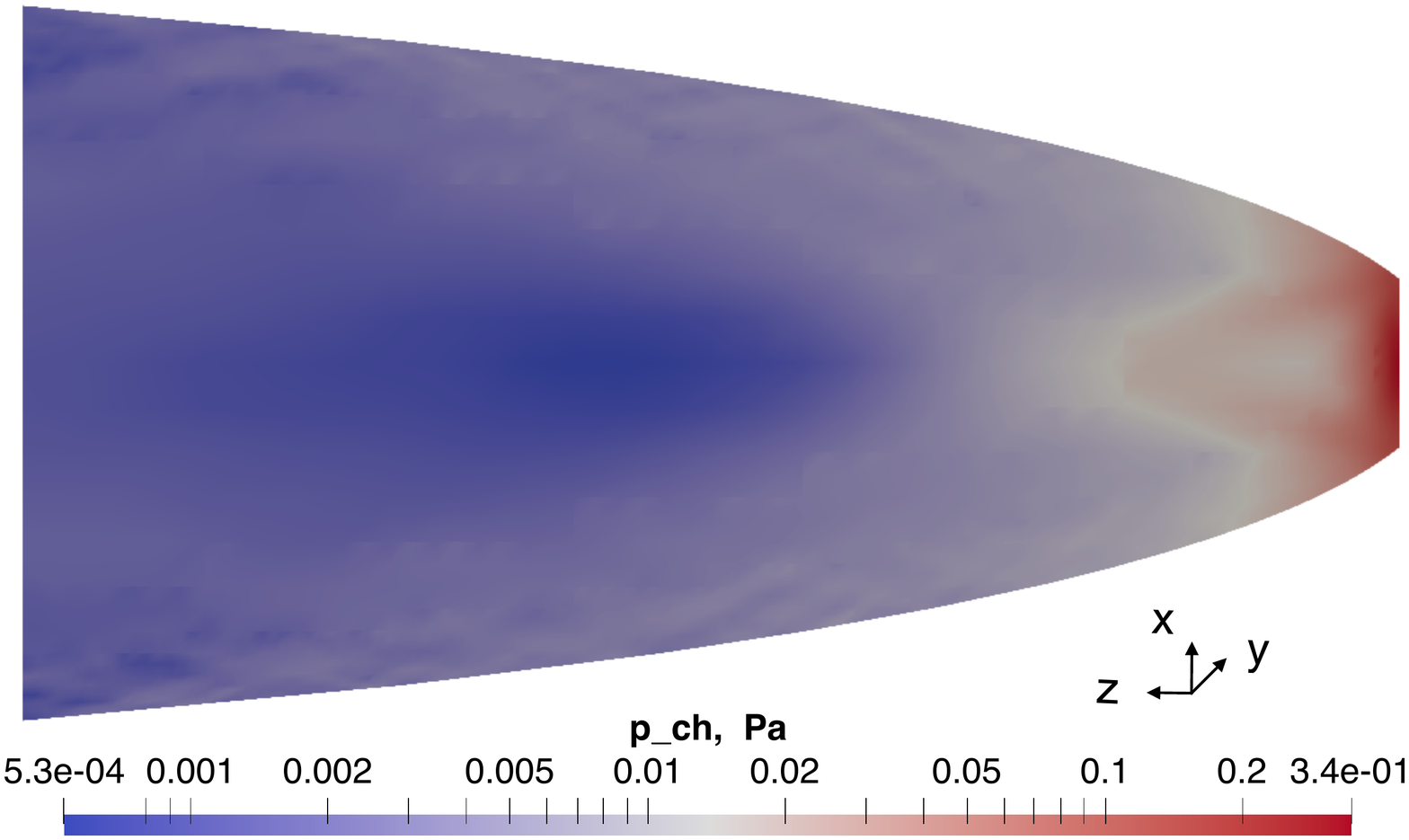}
 \caption{S-FT case pressure distribution, $h=\SI{150}{\kilo\meter}$}
 \label{c10-press}
\end{figure}

\begin{figure}[H]
 \centering
 \includegraphics[width=.85\textwidth, trim={3cm 0cm 3cm 0cm},clip]{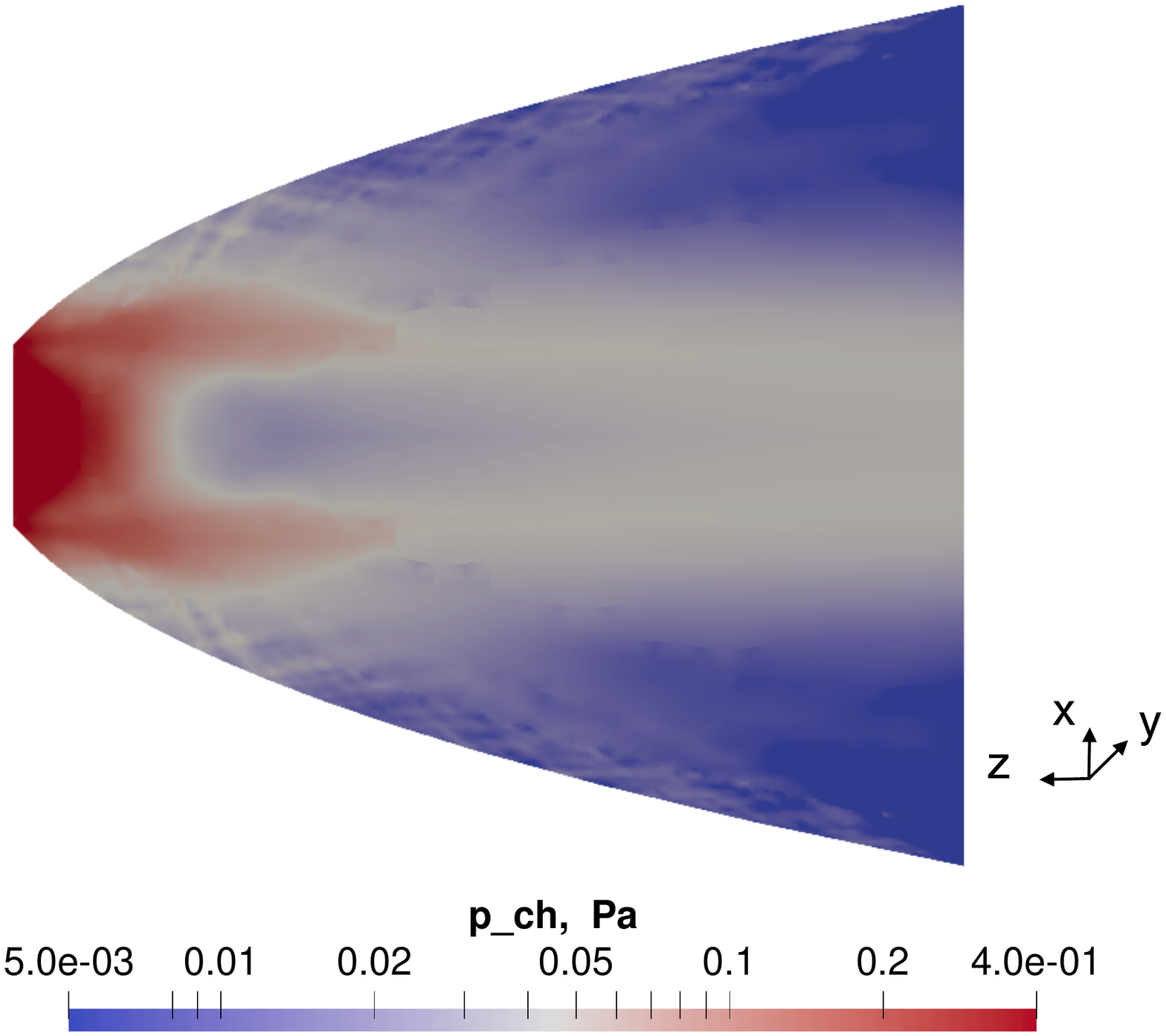}
 \caption{S-FE case pressure distribution, $h=\SI{150}{\kilo\meter}$}
 \label{c19_press}
\end{figure}

The specular intake improvement resulted in very efficient designs, especially the S-FT and the S-FT-D with a respective $\eta_c=0.943$ and $\eta_c=0.887$. Both are based on parabolic shapes with their focus located inside the thruster's discharge channel. The S-FT-D includes a front section of hexagonal ducts to reduce the backflow. In particular, the S-FT-D design resulted more sensitive to the misalignment with the flow $\alpha$ which sees a steeper drop of $\dot{m}_{thr}$ and $\eta_c$, see Fig.~\ref{fig:Ralt_pGrid}. The S-MS designs provide high $\dot{m}_{thr}$ but lower $\eta_c<0.334$ compared to the S-FT and S-FT-D designs. For comparison, the $p_{ch}$ distribution of the S-FT case and the S-FE are shown in Fig.~\ref{c10-press} and in Fig.~\ref{c19_press}. 
The different $p_{ch}$ distribution compared to diffuse-based intakes is due to the optical properties on which the particle collection is based. The S-FT, see Fig.~\ref{c10-press}, achieves $p_{ch} \sim \SI{0.3}{\pascal}$, while the S-FE achieves a larger pressure $p_{ch}>\SI{1}{\pascal}$ in the central region while also resulting in a greater larger-pressure region, see Fig.~\ref{c19_press}. However, it delivers lower $\eta_c=0.88$. It is also observed that the S-FT case provides an almost constant $\eta_c$ until $\alpha \sim \SI{10}{\degree}$. For $\alpha>\SI{10}{\degree}$ the impact on $\eta_c$ becomes larger. In particular, by increasing $\alpha=10 \rightarrow \SI{11}{\degree}$ there is a further decrease of $\eta_c$ by $\sim 10\%$. Such reduction continues until $\alpha=\SI{20}{\degree}$ at which $\eta_c \sim 0.13$. A $\alpha > \SI{15}{\degree}$, $\eta_c <  0.6$ and $\dot{m}_{thr} >\SI{0.1}{\milli\gram\per{\second}}$.

\subsection{Hybrid Intake Improvement}
The investigation on the combination of diffuse and specular reflecting materials on hybrid concepts differed from the expected result. It was initially thought that by having a diffuse chamber section, the intake performance would increase. However, lower $\eta_c$ compared to their pure specular counterpart are achieved. Furthermore, changing the percentage of diffuse versus specular scattering on the S-FT case provided insightful results for the understanding of the behaviour of the parabolic shape with different material properties. The $\eta_c$ is maintained $\eta_c>0.6$ for all cases with at least $70\%$ specular reflection. Although a specific relation of the $\sigma_B$ parameter to the scattering dynamics cannot yet be concluded, the results suggest an almost linear decrease of $\eta_c$ as the material becomes less specular and more diffuse, see Fig.~\ref{fcase10_R}. Such results are very valuable, as intake designs based on partial specular reflecting materials can provide $\eta_c > 0.6$. Additionally, the results show how intake performance could change due to possible material degradation over the mission lifetime, for example by the action of AO in VLEO~\cite{romanoacta}, and thus affecting the $\eta_c$.

\subsection{\textbf{Diffuse Intake}: Description, Performance, Render}
\begin{figure}[H]
    \centering
    \includegraphics[width=0.7\textwidth, trim={3cm 0cm 3cm 3cm},clip]{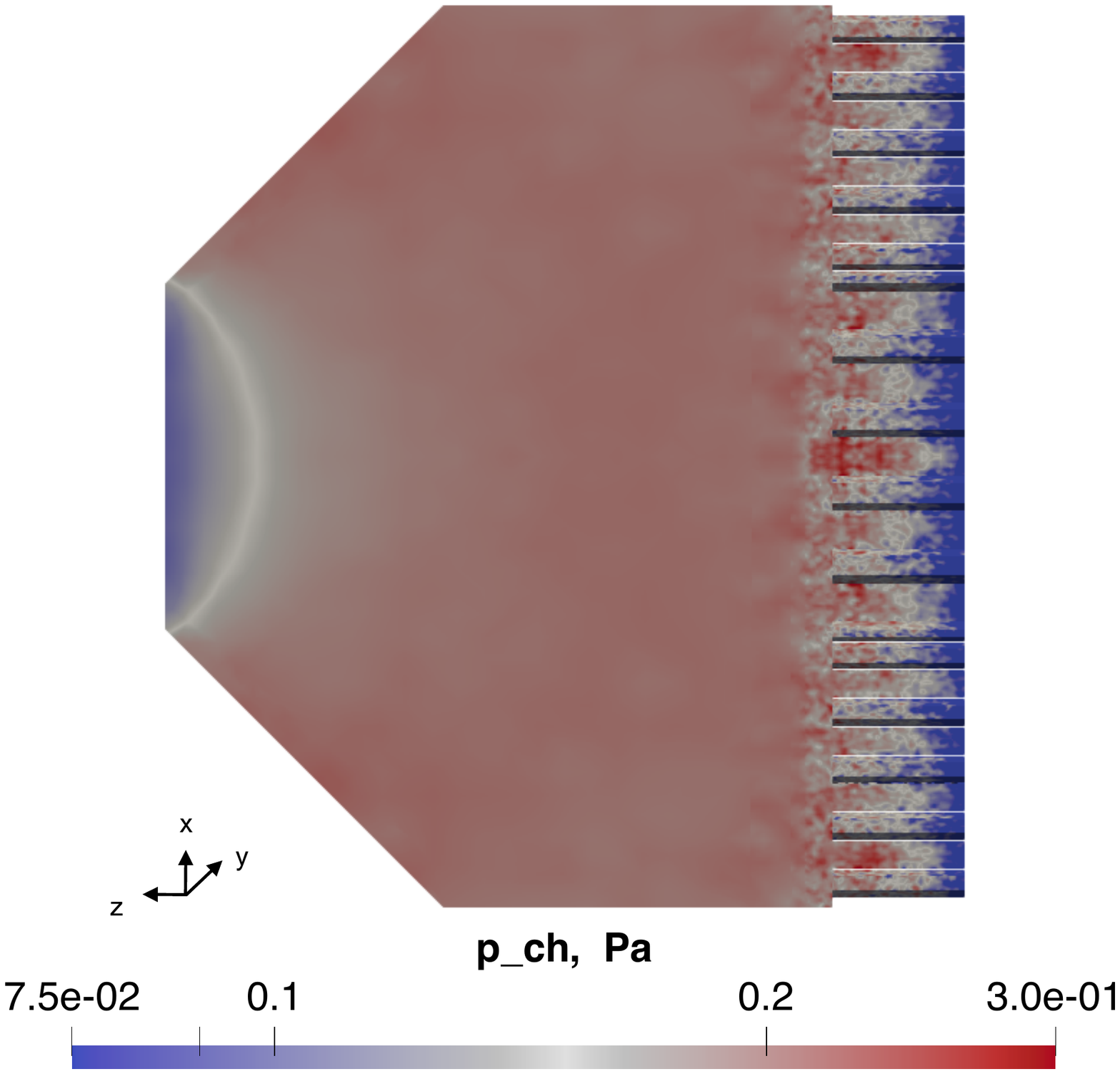}
    \caption{Diffuse Intake Pressure Distribution, $h=\SI{150}{\kilo\meter}$.}
    \label{c6_pres}
\end{figure}

In this section the final diffuse intake is presented. 
Based on the analysis of the intake concepts using diffuse scattering, the D-MDAR case is selected. Compared to the other best candidate, the D-MFR, provides a more homogenous pressure distribution, $p_{ch}\sim \SI{0.27}{\pascal}$, in front of the thruster's discharge channel as shown in Fig.~\ref{c6_pres}, while providing $\eta_c > 0.4$. 

The intake has an hexagonal shape and it features a front section of hexagonal ducts with different $AR$:
\begin{itemize}
\item Larger ducts in front of the thruster's discharge channel $\Rightarrow$ increase $A_{in}$;
\item Narrower ducts in the outer intake region $\Rightarrow$ reduce the backflow transmission probability, molecular trap operation.
\end{itemize} 
At the rear there is a conical section with a chamber angle of \SI{45}{\degree} terminating on the thruster's discharge channel diameter. 

Although the expected $\dot{m}_{thr}$ is low compared to the currently applied $\dot{m}_{thr}$ to the thruster's laboratory model~\cite{romanoiac4,romanosp2021}, 1) the thruster has still a large margin to be optimized in both size and operational conditions (therefore influencing intake geometry as well), 2) the input density is based on low solar activity, therefore for stronger solar activities, higher input density yields larger $\dot{m}_{thr}$ with limited variations of $\eta_c$. The pressure distribution within the diffuse intake is shown in Fig.~\ref{c6_pres}.

\begin{figure}[H]
    \centering
    \includegraphics[width=.55\textwidth, trim={13cm 6.5cm 16cm 5cm},clip]{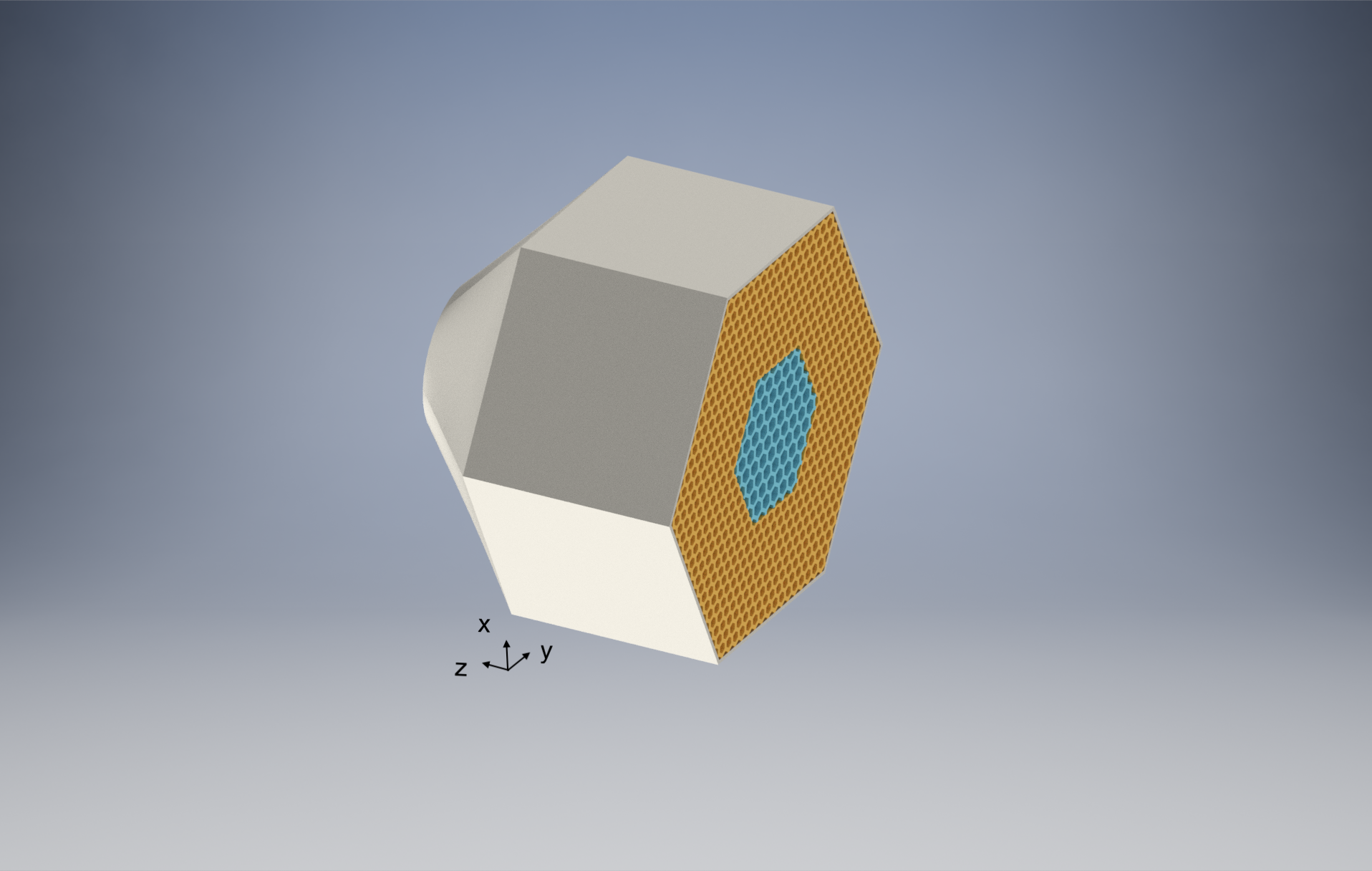}
    \caption{Diffuse Intake Render}
    \label{dif_intake_sch}
\end{figure}

In Tab.~\ref{tab:results-df-sum}, the respective intake performance in the selected VLEO altitude range are presented, while in Fig.~\ref{dif_intake_sch} the isometric render view of the diffuse intake is shown.

\begingroup
\setlength{\tabcolsep}{6pt} 
\renewcommand{\arraystretch}{1.5} 
\begin{table}[H]
\centering
\caption{Diffuse Intake Performance at Different Altitudes}
\label{tab:results-df-sum}
\begin{tabular}{cccccc}
\toprule
$h$ & $A_{in}$ & $\dot{N}_{in}$ & $\dot{N}_{out}$ & $\eta_{c}$ & $\dot{m}_{thr}$\\ 
$\SI{}{\kilo\meter}$ & $\SI{}{\meter^2}$ & $\SI{}{1/\second}$ & $\SI{}{1/\second}$ & - & $\SI{}{\milli\gram\per{\second}}$\\ 
\midrule
150 & 0.008 & 1.34E+18 & 6.11E+17 & 0.456 & 0.0240\\ 
180 & `` & 3.37E+17 & 1.49E+17 & 0.443 & 0.0053\\ 
200 & `` & 1.60E+17 & 7.14E+16 & 0.445 & 0.0023\\ 
220 & `` & 8.26E+16 & 3.51E+16 & 0.424 & 0.0011\\ 
250 & `` & 3.36E+16 & 1.40E+16 & 0.416 & 0.0004\\ 
\bottomrule
\end{tabular}
\end{table}
\endgroup

\subsection{\textbf{Specular Intake}: Description, Performance, Render}
In this section the final specular intake is presented. 
Based on the analysis based on specular scattering, the S-FT case is selected: a parabolic scoop with its focus inside the thruster discharge channel providing $\eta_c=0.94$, and $\dot{m}_{thr}=\SI{0.23}{\milli\gram\per{\second}}$ at $h=\SI{150}{\kilo\meter}$. Different simulation times, verifying runs, solar activity maximum-minimum, bow shock analysis, and different orbital altitudes that are not reported here for the sake of brevity all confirmed the same performance values. Such design outperforms the diffuse intake also in terms of sensitivity to flow misalignment leading to a limit of $\alpha=\SI{15}{\degree}$ resulting in $\eta_c<0.6$. The pressure achieved at the back of the intake is $p_{ch}=\SI{0.3}{\pascal}$.   
 
The S-FT case is selected as the best specular intake candidate as it provides $\eta_c = 0.94$ and $\dot{m}_{thr}\sim\SI{0.23}{\milli\gram\per{\second}}$. 

\begin{figure}[H]
    \centering
    \includegraphics[width=.8\textwidth, trim={9cm 2cm 8cm 2cm},clip]{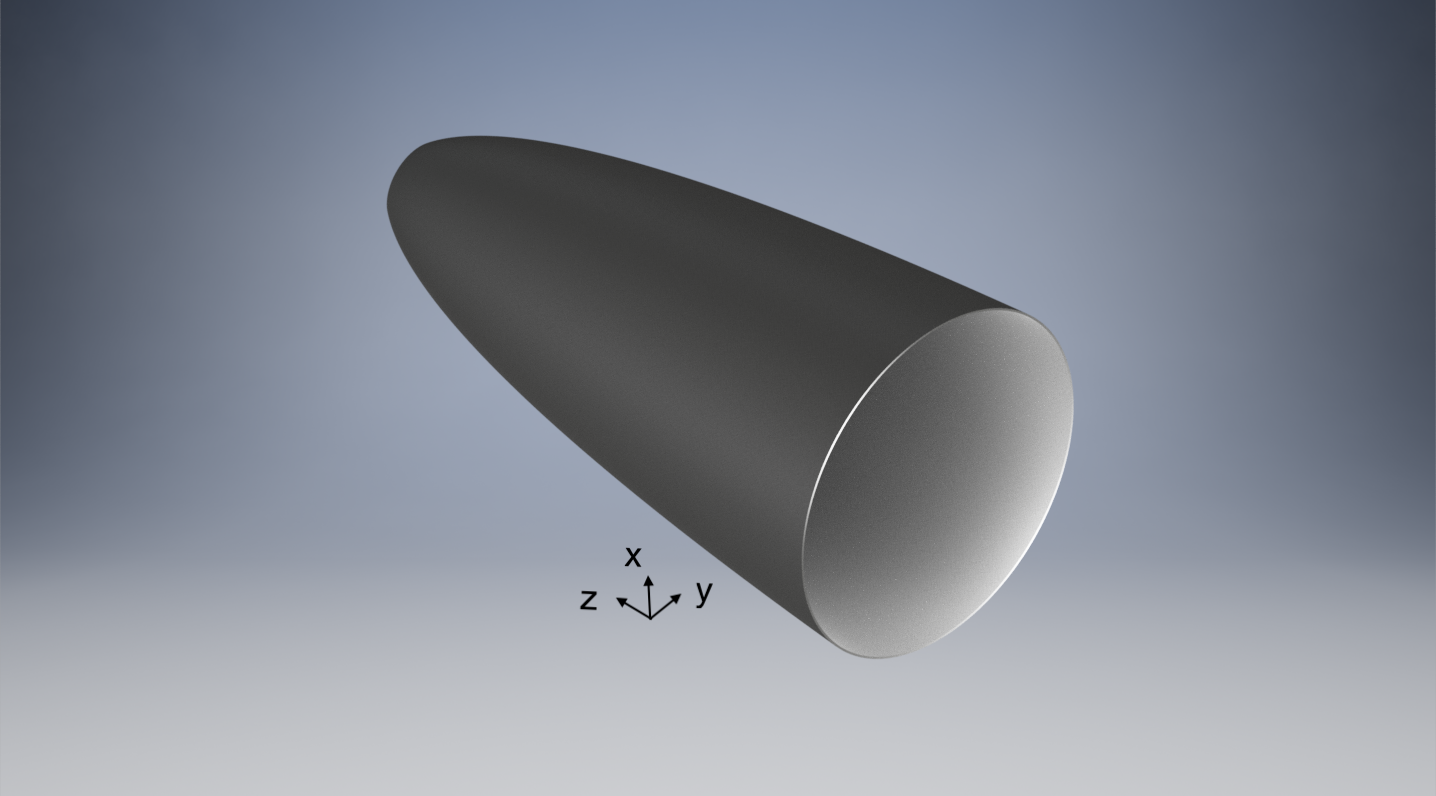}
    \caption{Specular Intake Render}
    \label{opt_intake_rend}
\end{figure}

Although it falls under the estimated value for the current laboratory model thruster's discharge channel operating pressure $p_{ch}=1 - \SI{7}{\pascal}$~\cite{romanoiac4,romanosp2021}, it has been experimentally shown that helicon plasma thrusters can operate as low as $p_{ch}=\SI{0.266}{\pascal}$~\cite{takahashi2019helicon}. The specular intake render is shown in Fig.~\ref{opt_intake_rend}, while the performances are shown in Tab.~\ref{tab:results-optimal}.

\begingroup
\setlength{\tabcolsep}{6pt} 
\renewcommand{\arraystretch}{1.5} 
\begin{table}[H]
\centering
\caption{Specular Intake Performance at Different Altitudes}
\label{tab:results-optimal}
\begin{tabular}{cccccc}
\toprule
$h$ & $A_{in}$ & $\dot{N}_{in}$ & $\dot{N}_{out}$ & $\eta_{c}$ & $\dot{m}_{thr}$\\ 
$\SI{}{\kilo\meter}$ & $\SI{}{\meter^2}$ & $\SI{}{1/\second}$ & $\SI{}{1/\second}$ & - & $\SI{}{\milli\gram\per{\second}}$\\ 
\midrule
150 & 0.019 & 6.33E+18 & 5.97E+18 & 0.943 & 0.232\\ 
180 & `` & 1.59E+18 & 1.49E+18 & 0.934 & 0.052\\ 
200 & `` & 7.58E+17 & 7.06E+17 & 0.930 & 0.023\\ 
220 & `` & 3.90E+17 & 3.62E+17 & 0.926 & 0.011\\ 
250 & `` & 1.59E+17 & 1.47E+17 & 0.922 & 0.004\\ 
\bottomrule
\end{tabular}
\end{table}
\endgroup

\subsection{ABEP System Performance Estimation}

To assess the ABEP system performance, a preliminary thrust vs drag analysis is hereby performed. 
The drag $F_{D}$ is calculated assuming the frontal area of the intake (specular and diffuse final designs) to be that of the spacecraft $A_f=A_{Intake}$, a drag coefficient $C_D =2.2$ 
is assumed~\cite{MOE2005793}. The SC velocity calculated as in Eq.~\ref{eq:v_orb}, and the respective density from the NRLMSISE-00 model, see Eq.~\ref{Eq:drag}.
\begin{equation}
F_{D}=\frac{1}{2} \rho(h) v(h)^2 A_{f} C_D
\label{Eq:drag}
\end{equation}
The thrust $F_{T}$ is calculated as in Eq.~\ref{Eq:thrust}.
\begin{equation}
F_{T}= \dot{m}_{thr} c_e
\label{Eq:thrust}
\end{equation}
Where $c_e$ is the exhaust velocity that, by applying the condition of full drag compensation $F_{D}/F_{T}=1$, the required $c_e$ can be extracted. Finally, as $c_e$ is linearly dependent with $v(h)^2$, and $v(h)$ changes only slightly in the considered VLEO $h$ range, $c_e=9.1-\SI{9.2}{\kilo\meter\per{\second}}$ for the ``Specular Intake", and $c_e=35.8-\SI{39.8}{\kilo\meter\per{\second}}$ for the ``Diffuse Intake". This highlights the greater effectiveness of the ``Specular Intake" compared to the ``Diffuse Intake", for example by maintaining the intake geometry but increasing $A_f$ such that $A_f>A_{in}$, to stay within the range achieved by the state-of-the-art conventional EP thrusters~\cite{EPReview}, an $A_f=\SI{0.1}{\meter^2}$ would require for the ``Specular Intake" $c_e\sim\SI{48}{\kilo\meter\per{\second}}$, while for the ``Diffuse Intake" $c_e\sim190-\SI{210}{\kilo\meter\per{\second}}$, way higher than what available with current EP technologies~\cite{EPReview}.
Finally, the ``Diffuse Intake" does not necessarily be discarded, as its performance can be still tuned for delivering higher $\dot{m}_{thr}$, and could benefit greatly from more accurate studies in low-drag SC design, and experimental results on the plasma thruster development. Moreover, it can be combined with a specular-based design to realize a hybrid intake.

\section{Conclusion}
This article provides with the intake working principle for an ABEP system, and describes the main equations to determine the main parameters of interest. Furthermore, a brief literature review of the currently designed intakes for ABEP systems, highlighting the respective performance is presented. GSI models are introduced and their importance in the intake design highlighted, along with the NRLMSISE-00 atmospheric model used to extrapolate VLEO conditions. Finally the numerical tool PICLas is introduced and utilized to simulate the designed intakes.
Hereby, it can be concluded that, concerning the diffuse-based intake design, $\eta_c$ is generally improved by the addition of small ducts of given length-to-radius ratios in the front of the intake, but their presence decreases $FR$, therefore $A_{in}$, leading to lower $\dot{m}_{thr}$, and $p_{ch}$, not to mention the $A_{Intake}>A_{in}$ that leads to increased aerodynamic drag. Larger ducts in front of the discharge channel can increase $A_{in}$ without increasing $A_{Intake}$. This, in turn, also leads to higher $\dot{m}_{thr}$, and consequently $\eta_c$, as the particles entering the central region right in front of the discharge channel cross section, and that are parallel to $z$, enter the discharge channel directly without impacting with the chamber walls. Indeed, by hypothetically removing the intake, therefore having $A_{in}=A_{out}$, the efficiency could be maximized leading to $\eta_c\sim 1$. 

An increment of $A_{in}$ can lead to higher $\dot{m}_{thr}$ and $p_{ch}$, however, a change of $A_{in}$ also modifies $A_{in}/A_{out}$ which, in turn, can decrease $\eta_c$. Moreover, an increase of $A_{out}$ generally leads to higher $\dot{m}_{thr}$ but, at the same time, lowers the pressure $p_{ch}$. Finally it is shown how a large area of uniform $p_{ch}$ can be built in front of the thruster's discharge channel. 

Concerning the general behaviour of the specular-based intake designs analysed here, $\eta_c$ is generally improved by shifting the position of the parabola focus to inside the thruster discharge channel. This also directly increases $\dot{m}_{thr}$ as most of the particles will be focused to that point, rather than to a point that is placed at the edge or before the edge of the thruster's discharge channel. The maximum pressure $p_{ch}$ is generally concentrated in the area around the focus point for the same reasons above, if a higher pressure before the thruster's discharge channel is desired, then the focus shall be moved earlier, at the edge of it. 
Compared to the diffuse case, the $FR$ is higher due to the absence of the small ducts in the front, leading to larger $A_{in}/A_{Intake}$ and, therefore, $\dot{m}_{thr}$.

The investigation of different intake geometries, based on diffuse and specular GSI, lead to two optimal intake designs for an ABEP system implementing the RF Helicon-based plasma thruster, one based on fully diffuse, and the other on fully specular reflections. The ``Specular Intake" offers the best performance in terms of $\eta_c$ with a final $\eta_c =0.94$ and a resulting $\dot{m}_{thr}\sim\SI{0.23}{\milli\gram\per{\second}}$ at $h=\SI{150}{\kilo\meter}$ for low solar activity, while requiring $c_e < \SI{9.2}{\kilo\meter\per{\second}}$ at VLEO altitudes for full drag compensation. Therefore, it provides a certain margin for compensating drag  of $A_f<\SI{0.1}{\square\meter}$ while maintaining $c_e <\SI{50}{\kilo\meter\per{\second}}$. Such design is also robust to misalignments with the flow compared to the diffuse intake. The performance drops at $\eta_c<0.6$ for $\alpha=\SI{15}{\degree}$, therefore small $\alpha$ variations can be tolerated during the mission, relaxing the requirement of the attitude control and determination system (ACDS). 
The ``Diffuse Intake", instead, provides $\eta_c=0.458$ for $A_{Intake}=\SI{0.004}{\square\meter}$ and $FR=52\%$. Such design is highly sensitive to the misalignment with the flow, highlighting a $-17\%$ drop in $\eta_c$ for $\alpha=\SI{5}{\degree}$. This results in a stronger requirement for the ACDS to maintain ABEP performance over the orbit. Moreover, hybrid intake taking advantage of both specular, and diffuse scattering properties are briefly investigated.
Furthermore, $\eta_c$ cannot be the only parameter used to compare two intakes that are based on different GSI properties, as design changes also influences areas and pressures, therefore, the large $\eta_c$ of the ``Specular Intake" certainly is advantageous but it always depends on is the thruster requirement, therefore making the ``Diffuse Intake" also a good solution for a given thruster. 
The use of the ``Specular Intake", moreover, directly drives the particles inside the discharge channel of the thruster, instead of building a slightly lower pressure compared and drive them due to diffusion based on the ``Diffuse Intake". Which one is best, finally depends on the thruster that is applied and its respective requirements.

Finally, in an ABEP-based missions, both the ACDS and ABEP systems must be designed to cope with such deviation. Moreover, the ABEP must be able to operate in a wide regime of density and composition as the residual atmosphere is not an uniform environment, as it depends not only on the altitude $h$, but also on the time, illumination, location, and solar activity. 

Further investigations are required for the $C_D$ and SC aerodynamic design, as well for AO resistant and drag-reducing materials~\cite{crisp2020-M,AOMAterials} to improve the modelling. Currently, a set of sub-scale prototypes are prepared for being tested in relevant VLEO AO flow conditions within the H2020 DISCOVERER project. The Rarefied Orbital Aerodynamics Research facility (ROAR) at The University of Manchester~\cite{ROAR} will be used to experimentally validate such designs under AO flows at VLEO conditions, as well as the future incoming results on aerodynamic SC control from the Satellite for Orbital Aerodynamics Research (SOAR)~\cite{SOAR}. Furthermore, the application of real material GSI properties to the simulations has to be applied for later being further iterated with the intake testing. This will serve to validate the models against experiments before building a flight version prototype. Moreover, the intake design work has to be strongly linked (and iterated along) with the thruster design work, to ensure their operation in a common range of pressures, mass flows, and efficiencies. Finally, an ABEP system has great potentials on Earth observation and telecommunications in VLEO over LEO~\cite{Crisp2020}, and a near future ABEP technological demonstration mission is needed. A GOCE-like mission~\cite{GOCE} could serve to experimentally validate the ABEP system. Novel mission types can arise as future exploration missions: long term high resolution data from Earth's magnetic field can be acquired with higher accuracy, as well as continuous measurements of the atmosphere and higher resolution imaging. A small platform equipped with both conventional and ABEP, would serve as technological demonstration. Furthermore, fully-ABEP-based SC can be applied also for exploration of celestials bodies other than Earth, such as Venus and Mars, and even further for the application to the gas giants and their moons.

\section*{Acknowledgements}
This project has received funding from the European Union's Horizon 2020 research and innovation program under grant agreement No.~737183. This reflects only the author's view and the European Commission is not responsible for any use that may be made of the information it contains.

\bibliographystyle{elsarticle-num}
\bibliography{bibliography}

\end{document}